\documentclass[a4paper,11pt,dvipsnames]{article}
\usepackage{amsmath,amssymb,amsfonts,amsthm,mathabx}
\usepackage[dvips]{graphics}
\usepackage{enumerate}
\usepackage{mathrsfs}
\usepackage{subfigure}
\usepackage{color}
\usepackage{hyperref}
\usepackage{float}
\usepackage[left]{lineno} %
\usepackage{multirow, makecell}
\usepackage{rotating}
\usepackage[T1]{fontenc}
\usepackage{lscape}

\usepackage[style=nature,url=false,isbn=false,%
maxbibnames=2]{biblatex}
\DeclareSymbolFont{AMSb}{U}{msb}{m}{n}

\usepackage{geometry}
\usepackage{url}

\DeclareMathOperator*{\argmax}{argmax}
\DeclareMathOperator{\NVI}{NVI}
\DeclareMathOperator{\Tr}{Tr}

\DeclareMathOperator{\NC}{NC}

\usepackage{amsfonts}
\usepackage{xcolor}
\usepackage{graphicx}

\title{\textbf{Multiscale mobility patterns and the restriction of human movement}}

\author{Juni Schindler
\thanks{Corresponding author: \href{mailto:juni.schindler19@imperial.ac.uk}{\small\texttt{juni.schindler19@imperial.ac.uk}}, ORCID iD: 0000-0002-8728-9286}
\and Jonathan Clarke
\and Mauricio Barahona\thanks{Corresponding author: \href{mailto:m.barahona@imperial.ac.uk}{\small\texttt{m.barahona@imperial.ac.uk}}, ORCID iD: 0000-0002-1089-5675} \\~\\
Department of Mathematics, Imperial College London, London, UK}

\date{}

\begin{document}\maketitle

\begin{abstract}\noindent
    From the perspective of human mobility, the COVID-19 pandemic constituted a natural experiment of enormous reach in space and time.
    Here, we analyse the inherent multiple scales of human mobility using Facebook Movement Maps collected before and during the first UK lockdown. First, we obtain the pre-lockdown UK mobility graph, and employ multiscale community detection to extract, in an unsupervised manner, a set of robust partitions into flow communities at different levels of coarseness. The partitions so obtained capture intrinsic mobility scales with better coverage than NUTS regions, which suffer from mismatches between human mobility and administrative divisions. Furthermore, the flow communities in the fine scale partition match well the UK Travel to Work Areas (TTWAs) but also capture mobility patterns beyond commuting to work. We also examine the evolution of mobility under lockdown, and show that mobility first reverted towards fine scale flow communities already found in the pre-lockdown data, and then expanded back towards coarser flow communities as restrictions were lifted. The improved coverage induced by lockdown is well captured by a linear decay shock model, which allows us to quantify regional differences both in the strength of the effect and the recovery time from the lockdown shock.
\end{abstract}


\section*{Introduction}

Spatiotemporal patterns of population mobility reveal important aspects of human geography, such as social and economic activity~\cite{simGreatCitiesLook2015}, the evolution of cities and economic areas~\cite{xuEmergenceUrbanGrowth2021}, the response to natural disasters~\cite{wangPatternsLimitationsUrban2016}, or the spread of human infectious diseases~\cite{brockmannHiddenGeometryComplex2013}. Whilst mobility patterns are linked to, and influenced by, both geographical and administrative boundaries~\cite{sobolevskyDelineatingGeographicalRegions2013}, they are also a direct reflection of social behaviour, and thus provide additional insights into the natural evolution of socio-economic interactions at the population level.
The increasing access to detailed and continuously updated mobility data sets from various sources (e.g., mobile devices~\cite{sobolevskyDelineatingGeographicalRegions2013}, GPS location traces~\cite{alessandrettiScalesHumanMobility2020}, Twitter data~\cite{wangPatternsLimitationsUrban2016}) opens up the opportunity to develop further quantitative approaches to harness the richness of such data~\cite{gonzalezUnderstandingIndividualHuman2008, songModellingScalingProperties2010, siminiUniversalModelMobility2012}.

Access to mobility data has recently become more widespread due to the COVID-19 pandemic, which prompted governments across the world to impose a range of restrictions on the daily activities and movements of their citizens~\cite{blavatnikschoolofgovernmentCoronavirusGovernmentResponse2021}. Such mobility data was of immediate use to refine and assess interventions targeting the spread of COVID-19~\cite{buckeeAggregatedMobilityData2020, changMobilityNetworkModeling2020,
    herrenDemocracyMobilityPreliminary2020, cintiaRelationshipHumanMobility2020,oliverMobilePhoneData2020, greaterlondonauthorityCoronavirusCOVID19Mobility2021}, and to evaluate the unequal effects of the pandemic across populations~\cite{changMobilityNetworkModeling2020}.
Yet, from the perspective of mobility, the pandemic also constituted a natural experiment of enormous reach in space and time that accelerated both the sharing of such data sets and the study of a severe mobility shock, in which human activities were curtailed to reduced areas for a sustained period~\cite{unwinStatelevelTrackingCOVID192020,nouvelletReductionMobilityCOVID192021,galeazziHumanMobilityResponse2021, bonaccorsiEconomicSocialConsequences2020, bonaccorsiSocioeconomicDifferencesPersistent2021, mollgaardUnderstandingComponentsMobility2022}.

An important aspect of mobility is the presence of inherent spatial and temporal scales as a result of, e.g., administrative divisions,  patterns of social interactions, jobs and occupations, as well as diverse means of transportation~\cite{simGreatCitiesLook2015}.
Recent work~\cite{alessandrettiScalesHumanMobility2020, arcauteHierarchiesDefinedHuman2020} has shown that this multiscale, nested structure of human activities contributes to the scale-free behaviour that had been previously found empirically~\cite{brockmannScalingLawsHuman2006,gonzalezUnderstandingIndividualHuman2008,songModellingScalingProperties2010}.

Here, we apply data-driven, unsupervised network methods to study the multiscale structure of UK mobility in data collected before and during the first COVID-19 lockdown. Data from user-enabled, anonymised `Facebook Movement maps' between UK locations~\cite{facebookdataforgoodDiseasePreventionMaps2020} is used to construct directed, weighted mobility graphs which are then analysed using unsupervised multiscale community detection~\cite{delvenneStabilityGraphCommunities2010, delvenneStabilityGraphPartition2013, lambiotteRandomWalksMarkov2014, bacikFlowBasedNetworkAnalysis2016}
to extract inherent flow communities at different levels of coarseness.
Hence, the inherent mobility scales emerge directly as robust flow communities in the data, obtained here through a scale selection algorithm.

Our results show that multiscale flow communities extracted from the baseline, pre-lockdown data broadly agree with the hierarchy of NUTS administrative regions (where NUTS stands for Nomenclature of Territorial Units for Statistics),
yet with distinctive features that result from commuting patterns cutting across administrative divisions. In addition, the flow communities at the fine scale match well Travel to Work Areas (TTWAs)~\cite{officefornationalstatisticsTravelWorkArea2016}, a geography of local labour markets computed by the UK Office for National Statistics from 2011 Census data on residency and place of work for workers aged 16+, but also capture human mobility patterns beyond commuting to work. 
We then quantify the extent to which mobility patterns under lockdown conform to the flow communities found in pre-lockdown data using data collected during the first UK COVID-19 lockdown (March--June 2020).  We find that the imposition of lockdown reverted mobility towards the local, fine scale pre-lockdown flow communities and, as restrictions were lifted, mobility patterns expanded back towards the coarser pre-lockdown flow communities, thus providing empirical evidence for the presence of a quasi-hierarchical intrinsic organisation of human mobility at different scales%
~\cite{alessandrettiScalesHumanMobility2020}.
Finally, we find regional differences in the response to the lockdown, both in the strength of the mobility contraction and the time scale of recovery towards pre-lockdown mobility levels.

\section*{Results}

We use mobility data provided by Facebook under the `Data for Good' programme~\cite{facebookdataforgoodDiseasePreventionMaps2020} to construct directed, weighted networks of human mobility in the UK.
The anonymised data sets (`Facebook Movement maps'~\cite{facebookdataforgoodDiseasePreventionMaps2020,maasFacebookDisasterMaps2019}), which are collected from user-enabled location tracking, quantify frequency of movement of individuals between locations over time, thus capturing temporal changes in population mobility before and during the COVID-19 pandemic~\cite{galeazziHumanMobilityResponse2021, bonaccorsiEconomicSocialConsequences2020}. For details of the network construction see Methods.

Our data covers mobility patterns in all four nations of the UK before, during and after the first nationwide COVID-19 lockdown, which was imposed on 24 March 2020 (see below for more details). In the following, we first analyse the pre-lockdown baseline mobility, from which we obtain intrinsic partitions at different scales, and then explore how the changes after lockdown mobility restrictions were imposed map onto those baseline scales.

\subsection*{The directed graph of baseline UK mobility: quasi-reversibility and commuting travel patterns}
\label{Sec:PropBaseline}

\begin{figure}[htb!]
    \centering
    \includegraphics[width=\textwidth]{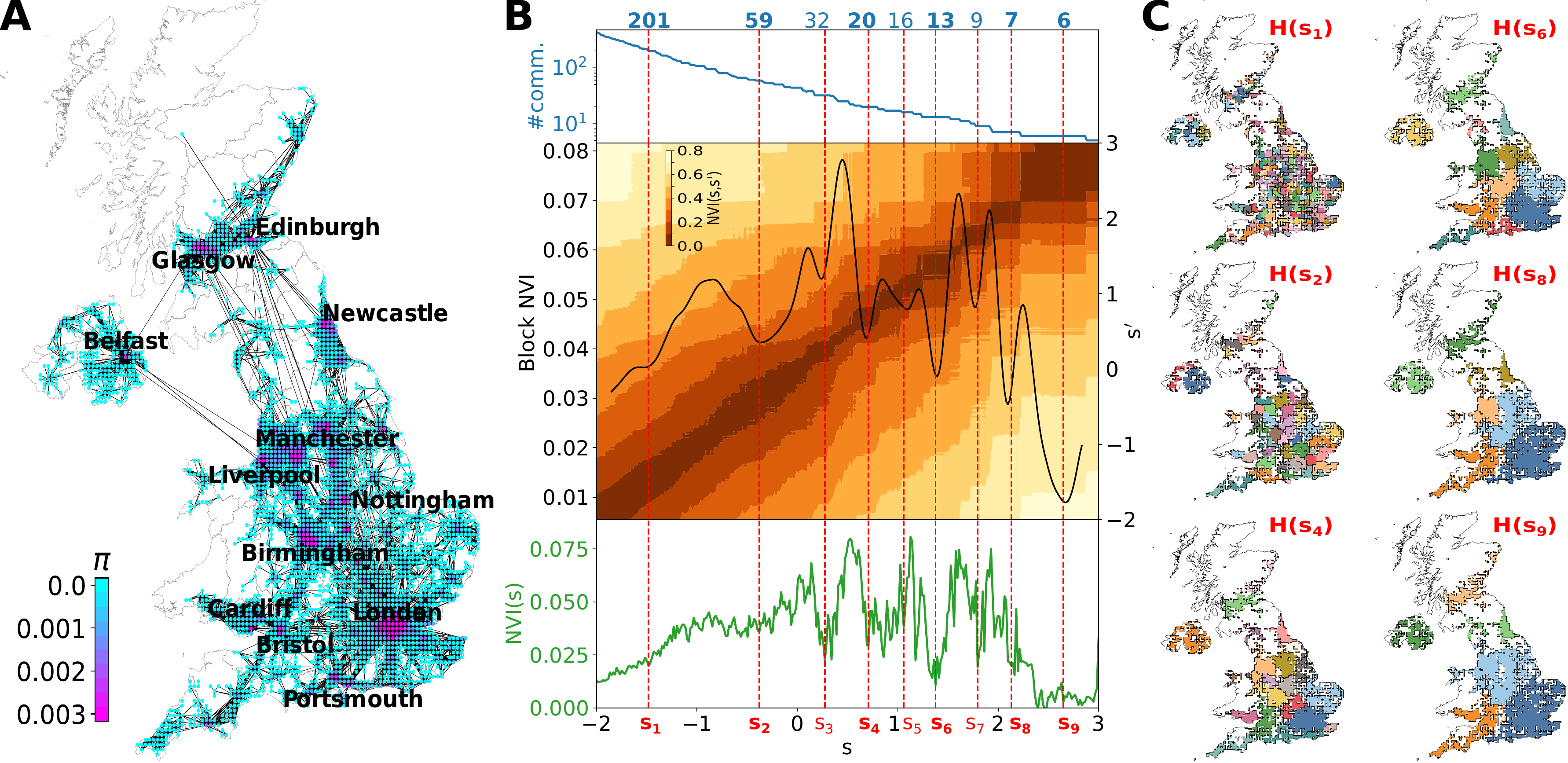}
    \caption{\textbf{Multiscale structure in the baseline mobility network.} \textbf{A} Using mobility data averaged over 45 days before 10 March 2020, we create a weighted directed graph ($N=3,125$ nodes, $E=34,224$ edges) with edge weights equal to the average daily number of trips between geographic tiles (nodes). The stationary distribution $\pi$~\eqref{eq:stationary} of the associated random walk indicates high centrality of urban areas.
        \textbf{B} Multiscale community detection on the baseline mobility network using Markov Stability analysis.
        We find optimised MS partitions that are robust both across scales (blocks of low $\NVI(s,s')$), and within scale (dips in $\NVI(s)$). To find robust optimal partitions, we first determine basins in Block NVI$(s)$ (the pooled diagonal of $\NVI(s,s')$)
        and then find the minima of $\NVI(s)$ for each basin (see Methods). This selection process leads to nine robust scales ($s_1, \dots, s_9$), from finer to coarser.
        \textbf{C} The graph partitions for the six scales with the lowest Block NVI %
        are plotted on the UK map, with different colours indicating different communities.
    }
    \label{fig:baseline_MS_analysis}
\end{figure}

Using pre-lockdown mobility data
(average of 45 days before 10 March 2020), we construct a strongly connected directed graph $G$ with weighted adjacency matrix $A\neq A^T$ (Fig.~\ref{fig:baseline_MS_analysis}\textbf{A}).
The $N=3,125$ nodes of this graph correspond to geographic tiles (width between 4.8--6.1 km, see Supplementary Fig.~\ref{S_fig:network_construction}\textbf{A}) and the directed edges have weights $A_{ij}$ corresponding to the average daily number of inter-tile trips from tile $i$ to $j$ (see Methods for the notion of `trip' within the Facebook data set, and some of its caveats). The total average number of daily inter-tile trips is $2,475,527$, as compared to $10,416,968$ intra-tile trips. The matrix $A$ is very sparse, with 99.7\% of its entries equal to zero, i.e., there are no direct trips registered between the overwhelming majority of tile pairs. Furthermore, the non-zero edge weights are highly heterogeneous, ranging from $1.4$ to $6,709$ daily trips (average $72.3$, coefficient of variation  $2.8$), underscoring the large variability in trip frequency across the UK.

To assess the directionality of the baseline network, we first compute the pairwise relative asymmetry (PRA) for each pair of tiles $ij$:
\begin{equation}\label{eq:PRA}
    0 \leq \text{PRA}_{ij} := \frac{|A_{ij}-A_{ji}|}{A_{ij}+A_{ji}} \leq 1,
\end{equation}
defined for pairs where $A_{ij}+A_{ji} > 0$.
The distribution of the $\text{PRA}_{ij}$ (Supplementary Fig.~\ref{S_fig:PRA_PDB_histogram}) shows that 25\% of the tile pairs have $\text{PRA} \geq 0.23$, a substantial asymmetry,
including 3,226 one-way connections (8.64\% of the total) with $\text{PRA}=1$.
It is thus helpful to use network analysis tools that can deal with directed graphs~\cite{beguerisse-diazInterestCommunitiesFlow2014}.

A natural strategy for the analysis of directed graphs is to employ a diffusive process on the graph to reveal important properties of the network, such as node centrality~\cite{brinAnatomyLargescaleHypertextual1998,arnaudonScaledependentMeasureNetwork2020} or graph substructures~\cite{lambiotteRandomWalksMarkov2014}, while respecting edge directionality.
We consider a discrete-time random walk on graph $G$ defined in terms of the $N \times N$ transition probability matrix $M$:
\begin{equation}\label{eq:Baseline_Transition_Prob}
    M := D_{\text{out}}^{+} A,
\end{equation}
where $D_{\text{out}}^{+}$ is the pseudo-inverse of $D_{\text{out}}=\text{diag}\left(\mathbf{d}_{\text{out}}\right)$, the diagonal matrix of out-strengths $\mathbf{d}_{\text{out}} = A \,  \mathbf{1}_N$. %
A key property of the random walk is its stationary distribution $\boldsymbol{\pi}$, a $1 \times N$ node vector defined through the equation
\begin{align}
    \label{eq:stationary}
    \boldsymbol{\pi} = \boldsymbol{\pi} M.
\end{align}
The component $\pi_i$ is a measure of centrality (or importance) of node $i$; a high value of $\pi_i$ means that the random walk on $G$ is expected to visit node $i$ often at stationarity~\cite{levinMarkovChainsMixing2009}.
This is equivalent to PageRank without teleportation
~\cite{perraSpectralCentralityMeasures2008}.
As expected, the centralities $\pi_i$ are highly correlated ($R^2 = 0.97$) with another node centrality measure, the out-strengths $d_{\text{out},i}$.
Interestingly, the centralities $\pi_i$ are also correlated with the intra-tile mobility %
($R^2 = 0.83$), a measure that is not used in the computation of $\boldsymbol{\pi}$, see Supplementary Fig.~\ref{S_fig:Pi_correlations} for details.
Therefore, urban areas display high centrality due to the concentration of human mobility in those areas (Fig.~\ref{fig:baseline_MS_analysis}\textbf{A}).

A random walk on a directed graph might still not display strong directionality at equilibrium. This is our finding here: the random walk defined by $M$ fulfils approximately the detailed balance condition:
\[ \Pi M \simeq \Pi M^T \quad \text{with} \quad
    \frac{\|\Pi M-\Pi M^T\|_{\text{F}}}{\|\Pi M\|_F} = 0.033,
\]
where $\Pi = \text{diag}(\boldsymbol{\pi})$ and $\|\cdot \|_F$ denotes the Frobenius norm
(see Fig.~\ref{S_fig:PRA_PDB_histogram} and~\cite{lambiotteRandomWalksMarkov2014,schaubMultiscaleDynamicalEmbeddings2019} for a more in-depth discussion).
The random walk for our mobility graph is therefore close to being time-reversible at equilibrium~\cite{levinMarkovChainsMixing2009}, so that the probability of following a particular
trajectory from node $i$ to $j$ is almost equal to the probability of going back on the same trajectory from $j$ to $i$. This property coincides with our intuition that most journeys in the mobility network are linked to commuting travel patterns.

\subsection*{Unsupervised community detection reveals intrinsic multiscale structure in the baseline mobility data}
\label{Sec:ApplicationMS}

To extract the inherent scales in the UK mobility data, we apply multiscale community detection to the baseline directed network.
We use Markov Stability (MS)~\cite{delvenneStabilityGraphCommunities2010}, a methodology that
reveals intrinsic, robust graph partitions across all scales
through a random walk Laplacian that simulates individual travel on the network based on the observed average daily trips. As random walkers explore the network, they remain contained within small subgraphs (communities) at shorter times, and then spill over onto larger communities at longer times. This definition of communities (and partitions) in terms of random walks makes the MS framework generally applicable to a wide range of network topologies, including directed networks, in contrast to standard hierarchical community detection algorithms~\cite{girvanCommunityStructureSocial2002,clausetFindingCommunityStructure2004,bonaldHierarchicalGraphClustering2018,}. MS uses an optimisation to identify graph communities in which the flow of random walkers is contained over extended periods, uncovering a sequence of robust graph partitions of increasing coarseness (regulated by the Markov scale $s$). In the context of our mobility network, this set of partitions captures intrinsic scales of human mobility present in the data. See Refs.~\cite{delvenneStabilityGraphCommunities2010, lambiotteLaplacianDynamicsMultiscale2009, lambiotteRandomWalksMarkov2014,schaubMarkovDynamicsZooming2012,delvenneStabilityGraphPartition2013, bacikFlowBasedNetworkAnalysis2016} and Methods for details of the methodology.

Fig.~\ref{fig:baseline_MS_analysis}\textbf{B} summarises the MS analysis for our network, which was carried out with the \texttt{PyGenStability} Python package~\cite{barahonaresearchPyGenStability2021}.
We find nine robust MS partitions $H(s_i)$ at different levels of resolution ($s_1, \ldots, s_9$) from fine to coarse, which comprise flow communities at different scales of human mobility (see Supplementary Table~\ref{S_tab:CommunityStats} and Supplementary Fig.~\ref{S_fig:MS_Sankey} for further statistics and visualisations). %
Fig.~\ref{fig:baseline_MS_analysis}\textbf{C} shows that these data-driven flow communities correspond to geographic areas, even though our data only contains relational mobility flows  without explicit geographic information. Furthermore, the nine partitions have a strong quasi-hierarchical structure, which is not imposed \textit{a priori} by our graph partitioning method (see Supplementary Fig.~\ref{S_fig:MS_Sankey} and \ref{S_fig:ConditionalEntropy}). The obtained partitions thus reflect an inherent multiscale structure in the patterns of UK human mobility.

\begin{figure}[htb!]
    \centering
    \includegraphics[width=\textwidth]{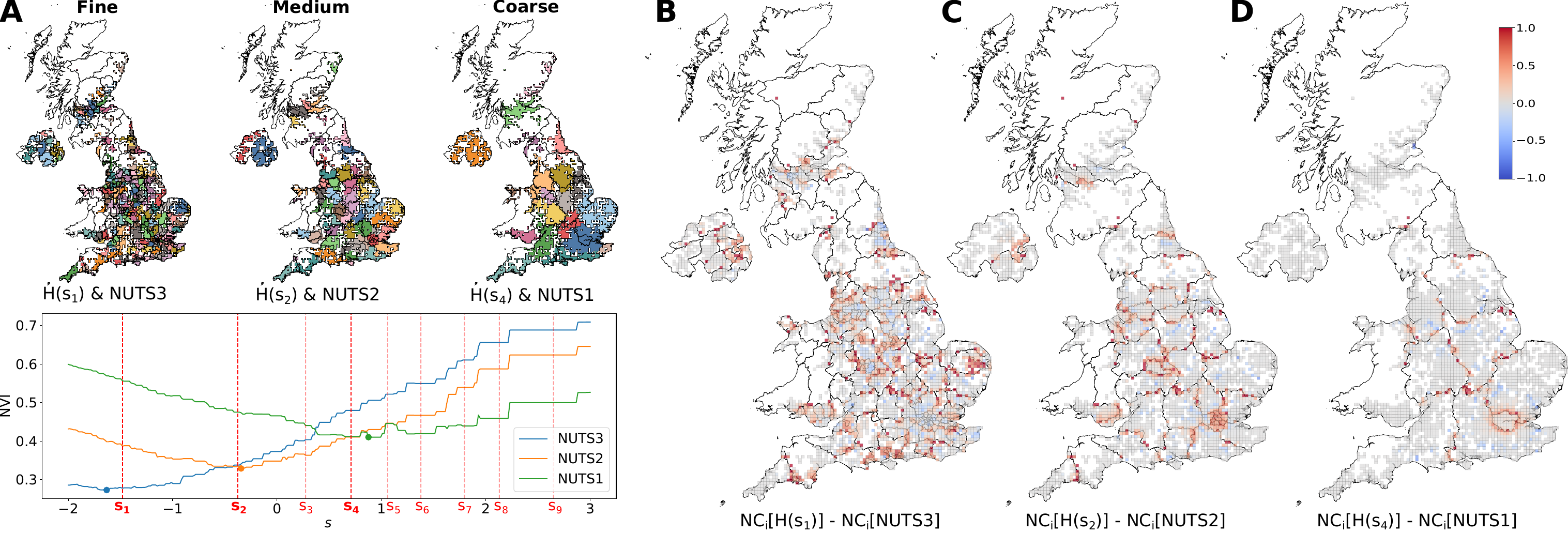}\caption{\textbf{\textit{A posteriori} comparison between Markov Stability (MS) partitions and NUTS regions.} \textbf{A} The MS partitions across all scales are compared to the three levels of administrative NUTS regions: NUTS3 (fine), NUTS2 (medium), and NUTS1 (coarse). As indicated by the minima of the NVI, the NUTS3 division is closely similar to $H(s_1)$; NUTS2 to $H(s_2)$; and NUTS3 to $H(s_4)$. The maps show NUTS regions (lines) and MS partitions (coloured communities) for fine, medium and coarse levels.
        \textbf{B-D} The MS partitions display improved average Nodal Containment (NC) relative to NUTS regions (Table~\ref{tab:TFC_IFC_Baseline}),
        and the node measure $\NC_i$~(Eq.~\ref{eq:NC_i}) shows the regional improvements where administrative divisions do not conform naturally to mobility patterns.
        Fine scale: $H(s_1)$ captures boundary areas fragmented under NUTS3 (e.g., Cornwall-Devon boundary, North Wales, among many others). Medium scale: $H(s_2)$ captures densely connected areas in Central and East London, fragmented under NUTS2, as well as the commuter belt in Birmingham. Coarse scale: $H(s_4)$ naturally captures Greater London's commuter belt that is excluded from the London NUTS1 region.
    }
    \label{fig:comparison_NVI_NC_i}
\end{figure}

\subsection*{Comparing the intrinsic mobility scales at baseline to administrative NUTS regions}
Next, we compare the MS partitions to NUTS regions,
administrative and geographic regions defined at three hierarchical levels: NUTS1 build upon NUTS2 in turn consisting of NUTS3 regions. In the UK, the 174 NUTS3 regions represent counties and groups of unitary authorities; the 40 NUTS2 regions are groups of counties; and the 12 NUTS1 regions correspond to England regions, plus Scotland, Wales and Northern Ireland as whole nations (see Supplementary Table~\ref{S_tab:NUTS_Stats} for further statistics). Our baseline data covers 170 NUTS3 regions, where the missing four are sparsely populated regions in the Scottish Highlands and Islands. 
The NUTS regions serve as a standard reference point for policy-making, and served to inform regionalised responses to COVID-19 in England (e.g., lockdowns in the North of England %
were applied to local authorities that form the NUTS2 region of Greater Manchester, Lancashire and West Yorkshire ~\cite{legislationukHealthProtectionCoronavirus2020c}). Comparing the data-driven MS partitions with NUTS regions is thus meant to explore to what degree administrative regions capture the patterns of mobility at the different scales, and potential mismatches thereof.

In Fig.~\ref{fig:comparison_NVI_NC_i} we use the Normalised Variation of Information (NVI)~\eqref{eq:NVI} to evaluate the similarity of each of the three NUTS levels to the MS partitions at all scales. The best match of each NUTS level (as given by the minimum of NVI) is close to one of the robust partitions: NUTS3 corresponds closely to $H(s_1)$; NUTS2 to $H(s_2)$; and NUTS1 to $H(s_4)$. Hence, these three MS partitions of the mobility network capture the fine, medium and coarse scales in the UK, yet with some significant deviations from the administrative NUTS divisions. For instance, Greater London is separated from the rest of the South East at the level of NUTS1 regions, whereas the whole South East of England forms one flow community in partition $H(s_4)$. Similarly, the south of Wales is connected strongly via flows to the South West of England in partition $H(s_4)$, which is not reflected in the NUTS1 regions. %
On the other hand, the correspondence between NUTS3 regions and the fine MS partition $H(s_1)$ is strongest (lower value of NVI), with fewer such discrepancies between administrative and flow communities.

To evaluate further how the MS partitions capture the patterns of mobility, we compute two measures: the Coverage $\mathcal{C}$~\eqref{eq:coverage_partition} (i.e., the ratio of mobility that remains within communities relative to the total mobility), and the average Nodal Containment $\NC$~\eqref{eq:NC_partition} (i.e., the ratio of the outflow from each node that remains within its community relative to the total outflow from that node, then averaged over all nodes). High values of these measures (normalised between 0 and 100\%) indicate that mobility flows are captured within the boundaries of the communities of the partition.

Table~\ref{tab:TFC_IFC_Baseline} shows that MS partitions are substantially better at reflecting baseline mobility than NUTS divisions since they have higher values for both average $\mathcal{C}$ and $\NC$ measures at all scales, and especially at the finer scales.
We have also evaluated both measures at a local level. Fig.~\ref{S_fig:coverage_NC_significance}\textbf{A} shows that the median of the Coverage of individual communities $\mathcal{C}_k$~\eqref{eq:coverage_k} is significantly higher for MS partitions (as compared to NUTS) for the fine and medium scales
($p<0.0001$, Mann-Whitney).
Fig.~\ref{S_fig:coverage_NC_significance}\textbf{B} shows that the median of the Nodal Containment of individual nodes $\NC_i$~\eqref{eq:NC_i}
is also significantly higher for MS partitions relative to NUTS regions at fine, medium and coarse levels ($p<0.001$, Mann-Whitney).
Indeed, the maps in Fig.~\ref{fig:comparison_NVI_NC_i}\textbf{B}-\textbf{D}
show that $\NC_i(\text{MS})>\NC_i(\text{NUTS})$ in regions where the administrative NUTS boundaries cut through conurbations or closely connected towns or cities. A prominent example is Greater London, where the NUTS2 regions split areas in Central and East London that are tightly linked and thus captured better by the medium MS partition $H(s_2)$,
and, similarly, the NUTS1 region of Greater London does not include its wider commuter belt that is naturally captured by the coarse MS partition $H(s_4)$.
Similar commuter belt effects are observed, e.g., on the medium level for Birmingham, and on the fine level for Plymouth, which has associated flows across the Cornwall-Devon boundary.

\begin{table}[htb!]
    \centering
    \begin{tabular}{ll||c|c|c}
&               & \textbf{Fine scale} & \textbf{Middle scale} & \textbf{Coarse scale} \\
&               & $H(s_1)$/NUTS3      & $H(s_2)$/NUTS2        & $H(s_4)$/NUTS1        \\
\hline \hline
\multirow{2}{4em}{\begin{turn}{00}$\mathcal{C}$ (\%) \end{turn}} & \textit{MS}   & 92.1                & 98.4                  & 99.7                  \\
& \textit{NUTS} & 90.1                & 95.2                  & 98.9                  \\
\hline
\multirow{2}{4em}{\begin{turn}{00} $\NC$ (\%)\end{turn}}                    &
\textit{MS}                                                                 & 86.6          & 95.5                & 98.1                                          \\
& \textit{NUTS} & 72.8                & 88.3                  & 95.8                  \\
\hline
    \end{tabular}
    \caption{\textbf{Containment of baseline mobility flows within MS partitions compared to corresponding NUTS regions.}
        MS partitions capture better the mobility patterns, as shown by higher values for both the average Coverage $\mathcal{C}$~\eqref{eq:coverage_partition} and the average Nodal Containment $\NC$~\eqref{eq:NC_partition}.
    }
    \label{tab:TFC_IFC_Baseline}
\end{table}

\subsection*{Comparing the fine mobility scale at baseline to labour-related Travel to Work Areas (TTWAs)}

We next compare the MS partitions to TTWAs, a different geography that divides the UK into 228 local labour markets computed from 2011 Census data recording place of residency and place of work ~\cite{officefornationalstatisticsTravelWorkArea2016}. Our baseline network has mobility data for 197 of the 228 TTWAs, with missing areas in rural areas of Scotland, Wales and the North of England (see Fig.~\ref{fig:comparison_TTWA}\textbf{A} and further statistics in Supplementary Table~\ref{S_tab:TTWA_Stats}).

\begin{figure}[htb!]
    \centering
    \includegraphics[width=0.5\textwidth]{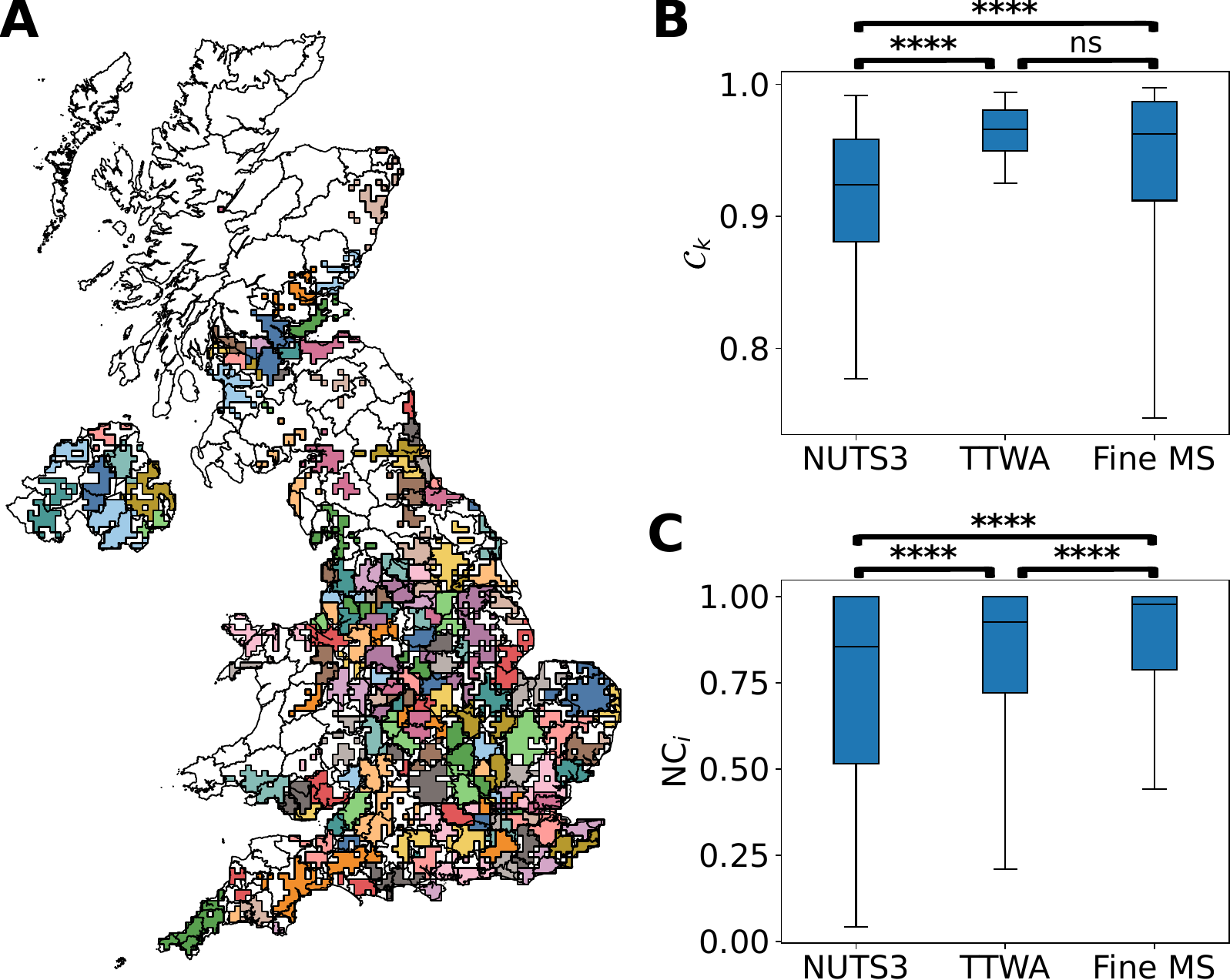}\caption{\textbf{\textit{A posteriori} comparison between the fine MS partition and TTWAs.} \textbf{A} The fine scale MS partition $H(s_1)$ is most similar to the division into TTWAs (Supplementary Fig.~\ref{S_fig:comparison_NVI_TTWA}) as shown by the overlayed map, where lines separate TTWAs and MS communities (fine scale) are indicated by different colours. Although well aligned, there are some discrepancies, e.g., $H(s_1)$ combines multiple TTWAs into a single community in Cornwall or Northern Ireland. \textbf{B,C} The flow coverage of fine MS partition is compared with NUTS3 and TTWAs: \textbf{B} community Coverage~\eqref{eq:coverage_k},  and \textbf{C} Nodal Containment~\eqref{eq:NC_i}. Both fine-MS and TTWA have significantly higher Coverage than NUTS3, and fine-MS has significantly higher Nodal Containment than both NUTS3 and TTWA. Statistical significance determined using the Mann-Whitney test (**** indicates $p<0.0001$). 
    }
    \label{fig:comparison_TTWA}
\end{figure}

The TTWAs are intended to reflect local labour markets, and the ensuing commuting between home and place of work, and are thus expected to be linked to small scales. Indeed, as measured by the NVI, the TTWA division is most similar to the NUTS3 level of all NUTS divisions (see Supplementary Information) and, consistently, most similar to 
the fine scale MS partition, $H(s_1)$.  Reassuringly, $H(s_1)$ is more similar to TTWA than to NUTS3 %
, since both the TTWA division and our MS partitions are data-driven with a basis in mobility patterns. Yet, there are local discrepancies, e.g., $H(s_1)$ combines multiple TTWAs into a single cluster in Cornwall or Northern Ireland, whereas the single TTWA in Greater London corresponds to several smaller communities in $H(s_1)$ ( Fig.~\ref{fig:comparison_TTWA}\textbf{A}).

As above, we evaluate how the TTWA division captures the patterns of mobility. We find that the Coverage $\mathcal{C}(\text{TTWA})=95.9\%$ is higher than for both NUTS3 and $H(s_1)$  (Table~\ref{tab:TFC_IFC_Baseline}), 
but the median of the Coverage of individual communities $\mathcal{C}_k$, a local measure of coverage, is not significantly higher for the TTWAs relatively to $H(s_1)$ (Fig.~\ref{fig:comparison_TTWA}\textbf{B}). Furthermore, the average Nodal Containment  $\NC(\text{TTWA})=81.3\%$ is lower than $H(s_1)$ (Table~\ref{tab:TFC_IFC_Baseline}), and its local version shows that the median of the Nodal Containment of individual nodes $\NC_i$ is significantly lower for TTWA ($p<0.0001$, Mann-Whitney, Fig.\ref{fig:comparison_TTWA}\textbf{C}). Hence the fine MS partition $H(s_1)$ captures better the mobility patterns in our baseline data than the TTWA division. This can be explained by potential changes in commuting patterns since the 2011 Census data on which the TTWAs are based, and by the fact that Facebook mobility data also includes trips for leisure, commercial and other activities beyond commuting to work.

\newpage \subsection*{The contraction of UK mobility under lockdown and its relation to the baseline mobility multiscale network}%

The first nationwide COVID lockdown in the UK was
imposed on 24 March 2020, instructing the British public to stay at home except for limited purposes.
Over the following months, restrictions were gradually eased to allow pupils to return to school (1 June 2020 in England but 17 August 2020 in Scotland), businesses to reopen (non-essential shops reopened on 13 June 2020 in England but 13 July in Scotland), and people to travel more freely for leisure purposes (13 May 2020 in England but 8 July 2020 in Scotland)~\cite{grewalVariationResponseCOVID192021}. %
We have analysed the response to these  restrictions using the time-dependent Facebook Movement maps~\cite{facebookdataforgoodDiseasePreventionMaps2020}  from 10 March 2020--18 July 2020 (131 days or 18 weeks). We construct mobility networks for each day, $G(d)$, and week, $G(w)$, defined on the same nodes (i.e., tiles) as the baseline network $G$ (see Methods).

\begin{figure}[htb!]
    \centering
    \includegraphics[width=0.8\textwidth]{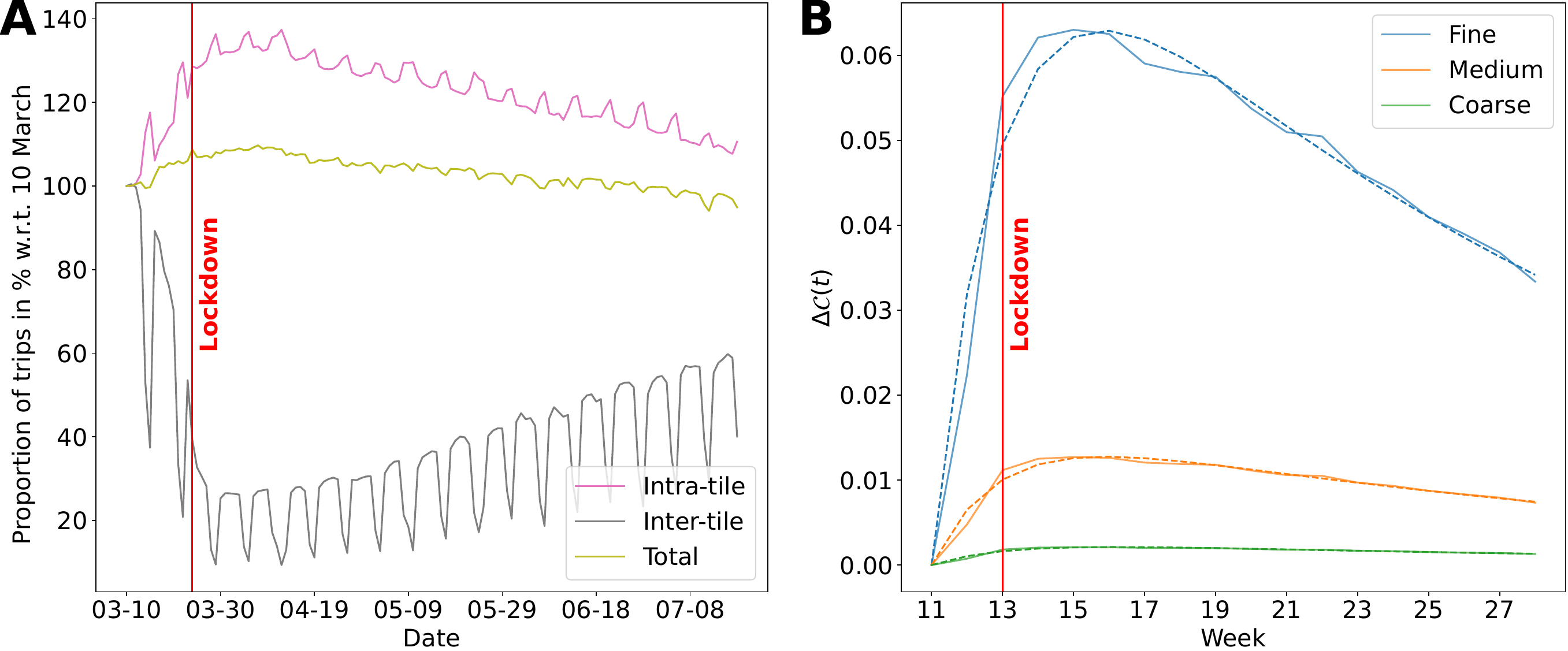}\caption{\textbf{Temporal response of mobility networks to lockdown restrictions.}
        The vertical red lines mark the official start of the first lockdown in the UK (24 March 2020). \textbf{A} Proportion (in \%) of the number of daily trips (intra-tile, inter-tile, total) with respect to the 10 March 2020. \textbf{B} Relative change of Coverage $\Delta \mathcal{C}$ of MS partitions (fine, medium and coarse scales)
        measured with respect to the first week of the study period (Week 11). $\Delta \mathcal{C}$  improves for all scales, especially for the finer scales. Solid lines correspond to the observed values; dashed lines are fits of the activation response function~\eqref{eq:response_solution}.
    }
    \label{fig:trips_coverage_response}
\end{figure}

Fig.~\ref{fig:trips_coverage_response}\textbf{A} shows the temporal change of the number of trips (intra-tile, inter-tile and total) relative to 10 March 2020, the first day of our study period. It is interesting to note that the decrease in mobility was already taking hold rapidly from 10 March, two weeks before the official enforcement of the lockdown. We find that, whilst the total number of trips remained largely unchanged
throughout the period, the number of inter-tile trips decreased sharply to $\sim$25\% of the initial value,
followed by a steady increase towards levels of around 50\% at the end of the study period in mid-July 2020.
Conversely, the number of intra-tile trips increased to a maximum of 130\% after lockdown
before decreasing steadily to around 105\% by mid-July 2020.
Therefore lockdown induced a redistribution from inter-tile to intra-tile trips
as a result of a reduction in commuting and long-distance travel, with mobility reverting to local neighbourhoods. %

The observed contraction of human mobility towards local neighbourhoods is consistent with the multiscale structure that was already present in the baseline mobility network pre-lockdown.
The Coverage $\mathcal{C}$ of all MS partitions increased over lockdown, with larger relative improvement for the finer scales (Fig.~\ref{fig:trips_coverage_response}\textbf{B} and Supplementary Fig.~\ref{S_fig:coverage_response_all_scales}).
The surge in Coverage induced by lockdown, which then decays towards its pre-lockdown value, can be modelled with a simple linear model under an external stimulus
$\alpha e^{-\lambda t}$.
The relative change from the initial value is then given by~\cite{beguerisse-diazLinearModelsActivation2016}:
\begin{align}\label{eq:response_solution}
    \Delta \mathcal{C}(t):=\frac{\mathcal{C}(t)-\mathcal{C}(t_0)}{\mathcal{C}(t_0)} = \frac{\alpha}{\beta-\lambda} \left (e^{-\lambda t} - e^{-\beta t} \right), %
\end{align}
from which we estimate the amplitude ($\alpha$) and characteristic time ($1/\lambda$) of the external stimulus, as well as the characteristic recovery time ($1/\beta$) of the system towards its pre-stimulus value  (see Methods for details).
Fig.~\ref{fig:trips_coverage_response}B shows the fits of $\Delta \mathcal{C}(t)$ (dashed lines) with estimated parameters in Table~\ref{Tab:Response_Parameters}. The fine MS partition exhibits the largest relative increase $\Delta \mathcal{C}(t)$ peaking at around 6\%.  The medium and coarse partitions peak at around $1\%$ and $0.1\%$, respectively. This is also captured by the values of $\alpha$, and indicates  that during lockdown people reverted to local mobility neighbourhoods already present in pre-lockdown patterns.
The adaptation to the new COVID situation and the pre-announcement of lockdown occurs quickly (over a characteristic time of $1/\lambda \sim 2\, \text{weeks}$), signifying that adoption was fast and was already in progress before the official start date of lockdown. Mobility patterns then returned towards pre-lockdown values over longer time scales $1/\beta$ between 16.4 weeks (fine scale) and 20.9 weeks (coarse scale) reflecting a slow re-adaptation following the new situation and loosening of restrictions.

\begin{table}[htb!]
    \centering
    \begin{tabular}[t]{l||c|c|c}
                              & $\alpha$ (95\% CI)      & $1/\beta$ (95\% CI) & $1/\lambda$ (95\% CI) \\
        \hline
        \hline
        \textbf{Fine scale}   & 0.042 (0.036--0.050)    & 16.4 (12.5--21.5)   & 2.0 (1.6--2.7)        \\
        \textbf{Medium scale} & 0.0086 (0.0074--0.0101) & 18.8 (14.6--24.3)   & 1.9 (1.5--2.5)        \\
        \textbf{Coarse scale} & 0.0014 (0.0012--0.0016) & 20.9 (15.7--28.0)   & 2.0 (1.6--2.6)        \\
        \hline
    \end{tabular}
    \caption{\textbf{Parameters of temporal response of $\Delta \mathcal{C}(t)$ for the MS partitions.} Estimated values and 95\% Confidence Intervals for the amplitude $\alpha$ and characteristic time $1/\lambda$ of the external stimulus, and the characteristic recovery time $1/\beta$ towards pre-stimulus values obtained from fitting the activation response function~\eqref{eq:response_solution} to the coverage values of the fine, medium and coarse MS partitions
        (see Fig.~\ref{fig:trips_coverage_response}\textbf{B}). Supplementary Table~\ref{S_tab:Coverage_All_Response_Parameters} provides all fitting parameters for all nine MS partitions.
    }
    \label{Tab:Response_Parameters}
\end{table}

\begin{figure}[htb!]
    \centering
    \includegraphics[width=0.6\textwidth]{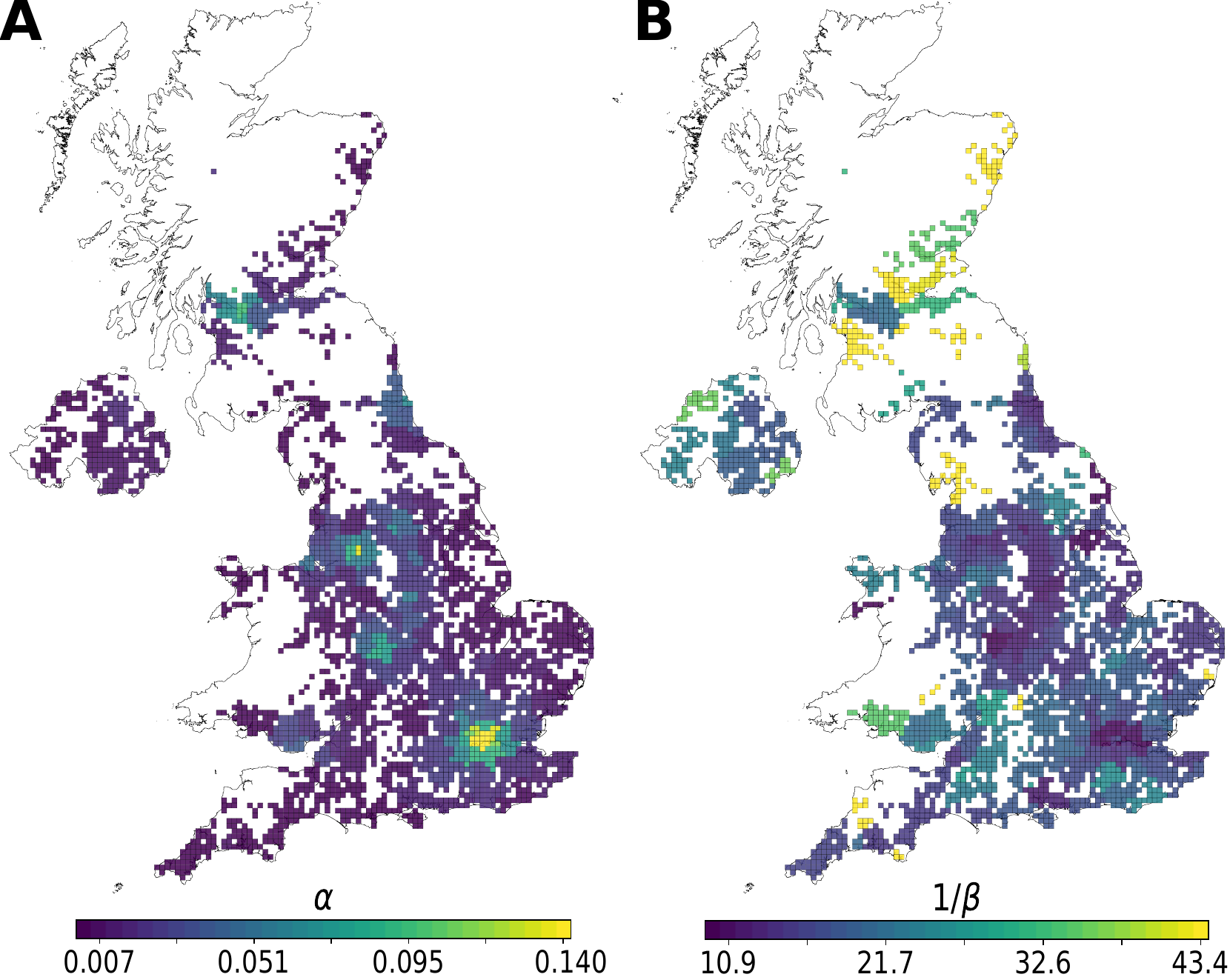}
    \caption{\textbf{Regional differences in the temporal response to the lockdown.} The maps show the fitting parameters of the activation response function for the weekly Coverage $\mathcal{C}_k$ of the communities in the finest MS partition. \textbf{A} The shock amplitude $\alpha$ is high for urban centres  (most notably London, Birmingham, Manchester and Glasgow) and low for rural areas. \textbf{B} The time scale of recovery $1/\beta$ is low for urban centres but high for rural areas, especially in Cornwall, Scotland or the Morecambe Bay area in North West England. Scotland shows longer recovery times due to different calendars for the lifting of restrictions relative to the rest of the UK.
    }
    \label{fig:respone_maps}
\end{figure}

To highlight the local differences in the temporal response to the lockdown, Fig.~\ref{fig:respone_maps} show the parameters of the temporal fits for the community coverages $\Delta \mathcal{C}_k$ for all the communities in the fine scale MS partition.
We observe that urban centres like London, Birmingham, Liverpool or Manchester experience the strongest changes in the fine scale Coverage (high values of $\alpha$) yet with faster recovery times (low values of $1/\beta$). Conversely, rural areas, which were already more constrained to local communities pre-lockdown, exhibit smaller but long-lived effects in the Coverage at the local level. Our method also captures divergent trends across the different nations of the UK. For example, Scottish regions show longer time scales of recovery than most regions in England, consistent with the fact that Scotland maintained more stringent lockdown restrictions for a longer time~\cite{grewalVariationResponseCOVID192021}, e.g., domestic travel restrictions were eliminated in Scotland only on 8 July 2020 and schools reopened on 17 August 2020, in contrast to 13 May and 1 June 2020 in England, respectively.

\section*{Discussion}\label{Sec:MS_Discussion}

Taking advantage of recent data availability, we have studied here the intrinsic multiscale structure of human mobility using, as a motivating example, UK data collected before and during the first COVID-19 lockdown.
Firstly, we generated a directed mobility graph from geospatial Facebook Movement maps collected before lockdown, and exploited multiscale graph clustering (MS) to identify inherent flow communities at different levels of resolution (or scales) in the baseline data in an unsupervised manner.
Three of the MS partitions so identified
are of similar granularity to the NUTS hierarchy, yet with improved mobility Coverage and Nodal Containment, also revealing areas of mismatch between human mobility and administrative divisions. Furthermore, we find that the fine MS partition, which captures local mobility in our data, shows high similarity to the division into TTWAs  obtained from Census residency and work location data to characterise local labour markets.

We then analysed spatiotemporal mobility data collected during the first UK COVID-19 lockdown.
We found
increased mobility Coverage for MS partitions, especially at the fine scale,
suggesting that the mobility contraction during this natural experiment reverted to scales already present in the pre-lockdown data.
Indeed, given that our MS communities are found through a random walk on a graph weighted by pre-lockdown trip frequency of natural mobility, the fine scales capture frequent trips that were not suppressed during lockdown, whereas the coarse scales are associated with less frequent trips over longer geographic distances for leisure or business.

The enhancement of Coverage induced by lockdown is well captured by a linear decay model, whose parameters allow us to quantify regional differences,
including differing trends across urban and rural areas and across the UK nations consistent with distinct lockdown regulations.

In this study, we have identified intrinsic communities at different scales extracted from a static network (our pre-lockdown baseline), and then studied how changes in mobility over the early months of the pandemic evolved relative to those inherent communities. The aim was to test the relevance of the inherent scales under this natural experiment by quantifying the extent to which mobility patterns conformed to communities derived from the baseline configuration. A complementary approach would be to instead obtain communities through additional analysis of the sequence of daily (or weekly) mobility networks using temporal community detection, e.g. via the recently proposed flow stability~\cite{bovetFlowStabilityDynamic2022}, an extension of MS to temporal networks. This would be an interesting direction for future research. 

Our study has several limitations.  Whilst the `Facebook Movement map' data is aggregated from 16 million UK Facebook users who enable location sharing (over 20\% of the total UK population), the observed sample might be biased and not representative of the general UK population~\cite{maasFacebookDisasterMaps2019}.
Furthermore, inter-tile flows with fewer than 10 trips within an 8-hour period are suppressed to prevent individual identification. In densely populated areas, our observed mobility data is thus more likely to be representative of human mobility, whereas this assumption is less likely to hold in rural areas, a limitation that can have an effect on comparisons with NUTS regions and TTWAs. However, such low-frequency connections account for a comparatively small number of the total trips and are not expected to affect the obtained MS partitions. This study also assumes rates of utilisation and activation of location sharing within the Facebook app remain constant, yet this limitation is mitigated by the derivation of flow partitions from average baseline data, rather than post-pandemic mobility information.

Our work contributes to the current interest in the study of intrinsic scales in human mobility~\cite{arcauteHierarchiesDefinedHuman2020}.
A recent study identified `spatial containers' from granular GPS traces ~\cite{alessandrettiScalesHumanMobility2020} organised in a nested hierarchy specific to each individual. Similarly, we also reveal a multiscale organisation of human mobility, but instead obtain a data-driven, unsupervised  quasi-hierarchical community structure at the population level.
Because our MS community detection is based on a diffusion on a mobility graph, the flow communities at different scales provide insights into the importance of physical and political geographies, and reveal the scales at which lockdown introduced frictions by restricting natural mobility.

\section*{Methods}

\subsection*{Mobility data and network construction}
\label{Sec:DataAggregation}

\paragraph{Data:}
Facebook Movement maps~\cite{facebookdataforgoodDiseasePreventionMaps2020,maasFacebookDisasterMaps2019} provide movement data between geographic tiles as codified by the \textit{Bing Maps Tile System}%
~\cite{schwartzBingMapsTile2018}. %
For the UK, there are 5,436 geographic tiles with widths between 4.8-6.1 km (see Supplementary Fig.~\ref{S_fig:network_construction}\textbf{A}). For users that enable location sharing, Facebook computes the dominant tiles, in which the user spends the most time over adjacent 8-hour time windows. The `trips' correspond to movements between dominant tiles across adjacent time windows. The dataset then provides the number of trips within each tile and to any other tile at intervals of 8 hours for all users. The data is anonymised by Facebook prior to release using proprietary aggregation methods, including the addition of small amounts of random noise, spatial smoothing, and dropping counts of less than 10 trips within an eight-hour period to avoid identifiability.  Our data further aggregates the three 8-hourly datasets for a given day. The data used in this study is the most comprehensive publicly available mobility dataset providing origin-destination data over time and covering the period of the COVID-19 pandemic. To our knowledge, no other dataset was available in the UK with better spatial and temporal resolution.

\paragraph{Network construction:}

Given a directed graph $G$, a weakly connected component (WCC) is a subgraph where each pair of nodes in the WCC is connected by an \emph{undirected} path. Similarly, in a strongly connected component (SCC) each pair of nodes is connected by a directed path~\cite{zweigNetworkAnalysisLiteracy2016}.
The largest weakly connected component is denoted as LWCC and the largest strongly connected component as LSCC.

As a baseline, we use pre-lockdown data consisting of mobility flows
averaged over the 45 days before 10 March 2020. To obtain the baseline network $G$, we remove the self-loops (i.e., we do not include intra-tile trips) and
we \emph{define} $G$ as the LSCC of the graph of flows.
As shown in Supplementary Fig.~\ref{S_fig:network_construction}, the LWCC and LSCC are similar and 98.8\% of the WCCs are singletons, hence the LSCC captures the large majority of relevant flows while simplifying the mathematical interpretation of the results.

We also use time series of mobility flows from 10 March 2020 to 18 July 2020 inclusive (131 days or 18 weeks)
to build daily mobility networks $G(d)$, $d = 1, ... , 131$ %
and weekly mobility networks $G(w)$, $w=1,...,18$,
(by averaging the daily networks over calendar weeks).
In all cases, the networks are defined on the same set of nodes as $G$, and we remove self-loops as above.

\subsection*{Multiscale community detection with Markov Stability analysis}

Here we provide a brief outline of the Markov Stability (MS) framework. For a fuller description,  see the Supplementary Information and in-depth treatments, including extensions to other types of graph processes, in Refs.~\cite{lambiotteLaplacianDynamicsMultiscale2009, delvenneStabilityGraphCommunities2010, delvenneStabilityGraphPartition2013, lambiotteRandomWalksMarkov2014}.

Consider a weighted and directed graph $G$ with adjacency matrix $A$. Let $L=I-D_{\text{out}}^{+} A$ denote the \textit{random walk Laplacian} matrix, where $I$ is the %
identity matrix, and $D_{\text{out}}^{+}$ denotes the pseudo-inverse of the diagonal out-degree matrix. The matrix $L$ defines a continuous-time Markov process on $G$ governed by the diffusive dynamics:
\begin{equation}\label{eq:diffusion}
    \frac{d \mathbf{p}}{dr} = -\mathbf{p}\:L,
\end{equation}
where $\mathbf{p}(r)$ is a $1 \times N$ node vector of probabilities, and $r$ is the Markov scale.
The solution to this equation is given by $\mathbf{p}(r)=\mathbf{p}(0) \exp(-Lr)$, and the matrix exponential defines transition probabilities of the Markov process
(see Supplementary Information).
This process converges to a stationary distribution $\boldsymbol{\pi}$ given by $\boldsymbol{\pi} L=0$.

The goal of MS is to obtain partitions of the graph into $c(r)$ communities such that the probability flow described by \eqref{eq:diffusion} is optimally contained within the communities as a function of $r$. MS solves this problem by maximising the following function:
\begin{align}
    \label{eq:MS_optimisation}
    H(r) =  \argmax_{H} \, \Tr\left[ H^T \left (\Pi \exp(-Lr) - \boldsymbol{\pi}^T \boldsymbol{\pi}\right ) H  \right]\, ,
\end{align}
where $\Pi = \text{diag}\left(\boldsymbol{\pi}\right)$, %
and the matrix $H(r)$ is a $N \times c(r)$ partition indicator matrix with $H(r)_{ij} = 1$ if $i$ is part of community $j$, and 0 otherwise.
We thus obtain a series of optimised partitions over the Markov scales described by the matrices $H(r)$.
The scales are more naturally described in log scale, so we \textit{redefine} the Markov scale as $s=\log_{10}(r)$.
The optimisation~\eqref{eq:MS_optimisation} is carried out using the Louvain algorithm~\cite{blondelFastUnfoldingCommunities2008} through the implementation in the \texttt{PyGenStability} python package~\cite{barahonaresearchPyGenStability2021}.

\subsubsection*{Comparing partitions with the Normalised Variation of Information }\label{Sec:NVI}

To assess the quality of the partitions
we use the Normalised Variation of Information (NVI), as a similarity measure for partitions~\cite{vinhInformationTheoreticMeasures2010,kraskovHierarchicalClusteringBased2003}.
Consider two partitions described by $H(s)$ and $H(s^{\prime})$ %
with potentially different numbers of communities. %
The NVI is defined as:
\begin{equation}\label{eq:NVI}
    0\le\NVI(s,s^{\prime}) := \frac{\text{VI}(s,s^{\prime})}{\mathcal{H}(s,s^{\prime})}\le 1, %
\end{equation}
where VI$(s,s')$ is the Variation of Information~\cite{meilaComparingClusteringsVariation2003} and $\mathcal{H}(s,s^{\prime})$ is the Mutual Information between $H(s)$ and $H(s')$.
The NVI is a metric and low values indicate a high similarity between the partitions~\cite{kraskovHierarchicalClusteringBased2003}.
Using NVI has the advantage of being a universal similarity metric~\cite{liSimilarityMetric2004}, i.e., if $H(s)$ and $H(s')$ are similar under any non-trivial metric, then they are also similar under NVI~\cite{kraskovHierarchicalClusteringBased2003}.

\subsubsection*{Scale selection algorithm}\label{Sec:ScaleSelection}

After obtaining optimised partitions $H(s)$ for a sequence of $m$ Markov scales $S=\{s_1,s_2,...,s_{m}\}$, we select partitions that describe the network structure robustly at different levels of resolution. Robust partitions are persistent across scales and reproducible under the non-convex Louvain optimisation for its particular scale~\cite{lambiotteRandomWalksMarkov2014}. We formalise these requirements using NVI as follows: (i) The persistence across scales is assessed by computing the pairwise NVI for partitions across different scales $s$ and $s'$ leading to a $m\times m$ symmetric matrix denoted by $\NVI(s,s')$, where regions of low values indicate high persistence across scales.
(ii) For each Markov scale $s$, the robustness is evaluated by repeating the Louvain optimisation (300 times in our study) with different random initialisation and computing the average pairwise NVI for the resulting ensemble of partitions, denoted by $\NVI(s)$ such that low values indicate strong reproducibility of the optimisation.

As an aid to scale selection, we propose here an algorithm that processes the information contained in $\NVI(s,s')$ and $\NVI(s)$ sequentially. First, we use tools from image processing to evaluate the block structure of the $\NVI(s,s')$ matrix and apply average-pooling~\cite{boureauTheoreticalAnalysisFeature2010} with a kernel of size $k$ (and padding) such that the pooled diagonal $\widehat{\NVI}(s)$ quantifies the average pair-wise similarity of all partitions corresponding to scales in the neighbourhood $\mathcal{B}_k(s)=\{u \in S: 0<|u-s|\le k\}$ of scale $s$.
We then compute the smoothed version of $\widehat{\NVI}(s)$, denoted Block NVI$(s)$.
Blocks of low values of $\NVI(s,s')$ correspond to basins around local minima of the Block NVI$(s)$.
We then obtain the minimum of $\NVI(s)$ for each basin, and determine those as the robust scales of the network.
Our scale selection algorithm is implemented in the \texttt{PyGenStability} package~\cite{barahonaresearchPyGenStability2021}.

\subsection*{Measures of flow containment: Coverage and Nodal Containment}\label{Sec:FlowCoverage}

Consider the adjacency matrix $A$ of the mobility graph $G$ and a $N\times c$ indicator matrix $H$ for a partition of $G$ into $c$ communities. Let us also define $\tilde{A}$, the adjacency matrix of the graph with self-loops that contains the intra-tile flows on the diagonal. Then $F=H^T \tilde{A} H$ is the $c \times c$ lumped adjacency matrix where the element $(H^T \tilde{A} H)_{kl}$ corresponds to the mobility flow from community $k$ to community $l$. The Coverage of community $k$, $\mathcal{C}_k(H)$, is defined as:
\begin{equation}\label{eq:coverage_k}
    0 \le \mathcal{C}_k(H) := (\hat{D}^+F)_{kk} %
    \le 1,
\end{equation}
where $\hat{D}^{+}$ is the pseudo-inverse of $\hat{D}=\text{diag}(\hat{\boldsymbol{d}})$ where $\boldsymbol{\hat{d}}=F\,  \mathbf{1}_c$.
$\mathcal{C}_k(H)$ can be interpreted as the probability of the lumped Markov process to remain in state $k$; hence high values of $\mathcal{C}_k(H)$ indicate that community $k$ covers well the flows emerging from the community.

The Coverage of a partition  $\mathcal{C}(H)$ is standard, and is defined as the ratio of flows contained within communities by the total amount of flow~\cite{fortunatoCommunityDetectionGraphs2010}. It is easy to see that this is given by the weighted average
\begin{equation}\label{eq:coverage_partition}
    0 \le \mathcal{C}(H) = \frac{\sum_k \boldsymbol{\hat{d}}_k \mathcal{C}_k(H)}{\sum_k \boldsymbol{\hat{d}}_k} %
    \le 1.
\end{equation}
High values of $\mathcal{C}(H)$ indicate that mobility flows are contained well within the communities of the partition and movement across different communities is limited.

The \textit{Nodal Containment} $\NC_i$ of node $i$ quantifies the proportion of flow emerging from $i$ that is contained within its community in a partition $H$:
\begin{equation}
    \label{eq:NC_i}
    0 \le \NC_i(H) := \frac{(AH)_{iC_i}}{d_i} \le 1,
\end{equation}
where $C_i$ is the community of node $i$ and $d_i=(A \,  \mathbf{1}_N)_i$.
Large values of $\NC_i$ indicate that the mobility flows emerging from  node $i$ are largely contained within its assigned community, indicating a good node assignment. Hence $\NC_i$
measures the containment of flows from a node-centred perspective.

To obtain a partition-level measure, we define the \textit{average Nodal Containment} $\NC(H)$:
\begin{equation}\label{eq:NC_partition}
    0 \le \NC(H) := \frac{1}{N}\sum_{i=1}^N\NC_i(H) \le 1,
\end{equation}
where $N$ is the number of nodes.

\subsection*{Response to an exponentially decaying shock}

The response of a variable $x(t)$ to a shock can be modelled as a linear ODE under a stimulus $R(t)$:
\begin{equation}\label{ODE_response}
    \frac{dx}{dt} = - \beta x + R(t), \quad x(0)=0
\end{equation}
where $1/\beta$ is the characteristic relaxation time of the system, and we assume here an exponentially decaying external stimulus $R(t):= \alpha e^{-\lambda t}$, with amplitude $\alpha \geq 0$ and characteristic decay time $1/\lambda$.
The solution of \eqref{ODE_response} is given by~\cite{beguerisse-diazLinearModelsActivation2016}:
\begin{align*}
    x(t) = \frac{\alpha}{\beta-\lambda} (e^{-\lambda t} - e^{-\beta t}).
\end{align*}

We use the Levenberg-Marquardt algorithm~\cite{moreLevenbergMarquardtAlgorithmImplementation1978} implemented in the LMFIT~\cite{newvilleLMFITNonLinearLeastSquare2014} python package to fit the activation response function $x(t)$ to a set of $n$ data points $(\tilde{t}_i,\tilde{x}_i)$ by minimising the sum of squares
\begin{equation}
    \chi^2 := \sum_{i=1}^n \left(x(\tilde{t}_i)-\tilde{x}_i\right)^2
\end{equation}
to determine parameter estimates $\hat{\alpha}$, $\hat{\beta}$, and $\hat{\lambda}$.
Confidence intervals (CIs) are obtained from an F-test~\cite{vugrinConfidenceRegionEstimation2007}.

\section*{Data and code availability}
Data used in this study was accessed through Facebook’s ‘Data for Good’ program: \url{https://dataforgood.facebook.com/dfg/tools/movement-maps}.
Shapefiles for the NUTS (2018) regions and TTWAs (2011) in the UK are available from the Open Geography Portal \url{https://geoportal.statistics.gov.uk/} under the Open Government Licence v.3.0 and contain OS data © Crown copyright and database right 2023.

We host data of the UK mobility networks alongside code to reproduce all results and figures in our study on GitHub: \url{https://github.com/barahona-research-group/MultiscaleMobilityPatterns}.

We use the \texttt{PyGenStability} python package~\cite{barahonaresearchPyGenStability2021} for Markov Stability analysis and scale selection. Code and documentation are hosted on GitHub under a GNU General Public License: \url{https://github.com/barahona-research-group/PyGenStability}.

\emergencystretch=1em
\setlength\bibitemsep{0pt}
\printbibliography

\section*{Acknowledgements}%
We thank Robert Peach and Michael Schaub for valuable discussions. We also thank Alex Pompe for insights regarding data collection. Finally, we thank our anonymous reviewers for their constructive feedback and helpful suggestions.

\section*{Funding}%
MB and JC acknowledge support from EPSRC grant EP/N014529/1 supporting the EPSRC Centre for Mathematics of Precision Healthcare. JC acknowledges support from the Wellcome Trust (215938/Z/19/Z). JS acknowledges support from the EPSRC (PhD studentship through the Department of Mathematics at Imperial College London).

\section*{Competing interests}
The authors declare that they have no competing interests.

\section*{Author contributions}
The computations in this study were performed by JS and JC. All authors contributed to the design of the work, and the writing of the manuscript.

\newpage

\renewcommand{\thepage}{S\arabic{page}} 
\renewcommand{\thesection}{S\arabic{section}}  
\renewcommand{\thetable}{S\arabic{table}}  
\renewcommand{\thefigure}{S\arabic{figure}}
\renewcommand{\theequation}{S.\arabic{equation}}

\appendix

\section{Multiscale community detection with Markov Stability analysis}

The directed graph of baseline mobility flows is analysed using Markov Stability (MS), a multiscale community detection framework that uses graph diffusion to detect communities in the network at multiple levels of resolution. MS is naturally applicable to directed graphs. 
We provide definitions and a summary of the formalism in the subsections below. %
For a fuller explanation of the ideas underpinning the method and several applications to social and biological networks, see~\cite{delvenneStabilityGraphCommunities2010, lambiotteLaplacianDynamicsMultiscale2009, delvenneStabilityGraphPartition2013,beguerisse-diazInterestCommunitiesFlow2014, lambiotteRandomWalksMarkov2014, bacikFlowBasedNetworkAnalysis2016,liuGraphbasedDataClustering2020}.

\subsection*{Diffusive processes on graphs} \label{Sec:diffusion_dynamics}
Consider a directed weighted graph $G$ with $N$ nodes and no self-loops.
Let $A \neq A^T$ be the $N \times N$ adjacency matrix of $G$ 
and $\mathbf{d}_{\text{out}} = A \,  \mathbf{1}_N$ be the vector of out-strengths, where $\mathbf{1}_N$ is the $N$-dimensional vector of ones.  Let us also define $D_{\text{out}}=\text{diag}\left(\mathbf{d}_\text{out}\right)$, the diagonal matrix containing the out-strengths on the diagonal. The transition probability matrix $M$ of a discrete-time random walk on $G$ is:
\begin{equation}\label{S_eq:transition_prob}
    M = D_{\text{out}}^{+} A,
\end{equation}
where $D_{\text{out}}^{+}$ denotes the pseudo-inverse of $D_{\text{out}}$.
The matrix $M$ defines a discrete-time Markov chain on the finite state space defined by the nodes of $G$~\cite{gallagerStochasticProcessesTheory2013}:
\begin{equation}\label{S_eq:random_walk} 
\mathbf{p}_{r+1}= \mathbf{p}_r \, M
\end{equation}
where $\mathbf{p}_r$ is a $1 \times N$ probability vector with components equal to the probability of the random walk hitting the respective  node at (discrete) time $r$. Clearly, $\mathbf{p}_r \cdot \mathbf{1}_N = 1, \, \forall r$, because $\mathbf{p}_r$ is a probability vector. While $\mathbf{p}_r$ defines the distribution of the Markov chain at time $r$, we denote by $X_r$ the random process that follows this distribution. %

A Markov chain on a finite state space is~\cite{gallagerStochasticProcessesTheory2013}:
\begin{itemize}
\item \emph{irreducible} when there is a positive probability to jump from an arbitrary state $i$ to another state $j$ in a finite number of steps; 
\item \emph{aperiodic} when the number of steps necessary to return from state $i$ to itself with positive probability has no fixed period; 
\item \emph{ergodic} when it is both irreducible and aperiodic.
\end{itemize}
A stationary distribution $\mathbf{\pi}$ of the Markov chain fulfills:
\begin{equation} \label{S_eq:stationary_dist_MC}
\boldsymbol{\pi} = \boldsymbol{\pi} \, M ,
\end{equation}
where $\boldsymbol{\pi}$ is a $1 \times N$ probability vector, which
corresponds to a dominant left eigenvector of $M$ with eigenvalue 1.
When the Markov chain is ergodic, it has a unique stationary distribution $\boldsymbol{\pi}$.

The random walk defined by $M$ is said to have the detailed-balance property~\cite{levinMarkovChainsMixing2009} if its stationary distribution $\boldsymbol{\pi}$ fulfills
\begin{equation}\label{S_eq:detailed_balance}
    \pi_i M_{ij} = \pi_j M_{ji}, \quad \forall  i,j=1,...,N
\end{equation}
which can be rewritten as a matrix equation 
\begin{equation}\label{S_eq:detailed_balance_matrix}
    \Pi M = M^T \Pi,
\end{equation}
where $\Pi = \text{diag}\left(\boldsymbol{\pi}\right)$.

A Markov chain $X_r$ with stationary distribution $\boldsymbol{\pi}$ that fulfills the detailed balance equation \eqref{S_eq:detailed_balance} is called reversible~\cite{levinMarkovChainsMixing2009}
since the distribution of the Markov chain up to each time $r$ is equal to the distribution of the time-reversed Markov chain, i.e.
\begin{equation}
     (X_0,X_1,...,X_r) \overset{d}{=} (X_r,X_{r-1},...,X_0),
\end{equation}
when $X_0$ is distributed according to the stationary distribution $\boldsymbol{\pi}$. Here $\overset{d}{=}$ denotes that the two random variables $(X_0,X_1,...,X_r)$ and $(X_r,X_{r-1},...,X_0)$ have the same distribution functions.

When, as in our case, the transition matrix $M$ is obtained from the adjacency matrix of a graph, the above conditions for the Markov chain can be translated into graph properties. In particular, it is well known that a necessary and sufficient condition for the irreducibility of the Markov chain defined by a graph-based transition matrix $M$~\eqref{S_eq:transition_prob} is that the graph $G$ is strongly connected~\cite{schaubMultiscaleDynamicalEmbeddings2019}.
The assumption of aperiodicity is harder to interpret but requires that the graph is non-bipartite. Under these conditions, the Markov chain defined by 
equation~\eqref{S_eq:transition_prob} is ergodic, and its unique stationary distribution $\boldsymbol{\pi}$ is related to PageRank without teleportation
~\cite{perraSpectralCentralityMeasures2008}.

There are continuous-time processes associated with the random walk~\eqref{S_eq:random_walk} (see Refs.~\cite{delvenneStabilityGraphPartition2013,  lambiotteRandomWalksMarkov2014} for a detailed discussion). In particular, let us define the rate matrix 
\begin{equation}\label{S_eq:QMatrix}
Q = M - I, 
\end{equation}
where $I$ is the $N \times N$ identity matrix. %
Note that $L=-Q$ is the so-called \emph{random walk Laplacian}. 
We can then define the continuous-time Markov process $X(r)$ with semi-group $P(r)$ governed by the forward Kolmogorov equation
\begin{equation}\label{ForwardKolmogorov}
    \frac{dP}{dr} = P\:Q,
\end{equation}
which has the solution
\begin{equation}\label{S_eq:MarkovSemigroup}
    P(r) = e^{rQ}.
\end{equation}
When the Markov process is ergodic, %
$P(r)$ converges in total variation to its unique stationary distribution $\boldsymbol{\pi}$, defined by Eq.~\eqref{S_eq:stationary_dist_MC}, which also fulfills $\boldsymbol{\pi} L=0$~\cite{scheutzowConvergenceMarkovChain2021}. 

From the theory of Markov processes~\cite{kuntzMarkovChainsRevisited2020}, it is known that for all $r>0$, the jump times 
\begin{equation}
\label{S_eq:jump_times}
    \tau = \inf\{ r>0: X(r) \neq X(0)\}
\end{equation}
are distributed according to
\begin{equation}
\label{S_eq:jump_times_distribution}
    \mathbb{P}[\tau \ge r] = e^{-q_i r}
\end{equation}
with jump rates $q_i = - Q_{ii}$. Furthermore, it is known that for all $
r>0$
\begin{equation}
    \mathbb{P}_i\left[\tau \ge r \; \text{and} \; \exists \; l=l(\omega)>0: X(u)=j \; \forall \; (\tau,\tau+l)\right] = e^{-q_i r} \frac{q_{ij}}{q_i}.
\end{equation}
with $q_{ij} = Q_{ij}$. Hence, the $ij$ element of the jump matrix is given by $q_{ij}/q_i$ and contains the jump probability of the Markov process $X(r)$ from state $i$ to $j$. Clearly, when $G$ has no self-loops, $q_i = - Q_{ii} = 1, \, \forall i$. Consequently, the jump times~\eqref{S_eq:jump_times} are exponentially distributed with rate 1, and  $M$ itself is the jump matrix so that
the jump probabilities of $X(r)$ are governed by $M$, as $q_{ij} = M_{ij}$ for $i \neq j$. %
These results show that the jump probabilities of the diffusion process are determined by the edge weights in the graph, whereas the self-loops of the graph afford the flexibility to choose different jump times reflecting distinct modelling assumptions~\cite{lambiotteRandomWalksMarkov2014}.

\subsection*{Markov Stability as a cost function for clustering algorithms}

The dynamics of the Markov chain with transition matrix $M$ defined on the nodes of the graph  can be exploited to get insights into the properties of the graph $G$ itself. 

Let $P(r)$ be the semi-group of a Markov process as defined in~\eqref{S_eq:MarkovSemigroup} with stationary distribution $\boldsymbol{\pi}$ on a graph with adjacency matrix $A$. Following~\cite{lambiotteRandomWalksMarkov2014}, each partition of the graph into $c$ communities corresponds to a $N \times c$ indicator matrix $H$ 
where $H_{ij} = 1$ if node $i$ is part of community $j$ and  $H_{ij} = 0$ otherwise. Define now the clustered autocovariance matrix for partition $H$ as
\begin{equation}\label{S_eq:autocovariance_matrix}
    K_r(H) = H^T \left[\Pi P(r) - \boldsymbol{\pi}^T \boldsymbol{\pi} \right] H.
\end{equation}
The diagonal elements $K_r(H)_{ii}$ correspond to the probabilities that the Markov process starting in one community $i$ does not leave the community up to time $r$, whereas the off-diagonal elements correspond to the probabilities that the process has left the community in which it started by time $r$. It is important to remark that $r$ is an intrinsic time of the Markov process that is used to explore the graph structure and is clearly distinct from the physical time of some applications, e.g., the days in the mobility data. To avoid confusion, it is customary in Markov Stability analysis to refer to $r$ and/or $s=\log_{10}(r)$ as the Markov \emph{scale}. Following these observations, we define the Markov Stability of a partition $H$ by
\begin{equation}\label{S_eq:MarkovStability}
    \mathcal{R}_r(H) = \min_{0\le l\le r} \text{Tr}\left[K_l(H)\right]
    \approx \text{Tr}\left[K_r(H)\right].
\end{equation}
 The approximation in~\eqref{S_eq:MarkovStability} is supported by numerical simulations that suggest that $\text{Tr}\left[K_r(H)\right]$ is monotonically decreasing in $r$.
The Markov Stability $\mathcal{R}_r(H)$ is thus a dynamical quality measure of the partition for each Markov scale $r$ which can be maximised to determine optimal partitions for a given graph and each scale of the associated Markov process. 

The objective is therefore to find a partition $H(s)$ that maximises Markov Stability up to a time horizon (scale) $s$ for the Markov process on the graph:
\begin{equation}
    \mathcal{R}_s\left(H(s)\right) = \max_H \mathcal{R}_s(H).
\end{equation}
The optimisation of this cost function is known to be NP-hard. However, Markov Stability can be re-expressed in terms of a generalised modularity~\cite{newmanModularityCommunityStructure2006} for another directed graph~\cite{lambiotteRandomWalksMarkov2014}, thus allowing the use of efficient algorithms developed for modularity optimisation, e.g., the Louvain algorithm~\cite{blondelFastUnfoldingCommunities2008}.

Briefly, the Louvain algorithm starts by assigning each node to a different community, followed by two steps. In the first step, one loops over the nodes in a random order and a node $i$ is added to a neighbouring community if the modularity increases. After no further increase is possible, a new network is generated with the communities as `super-nodes' and aggregated edges. The two steps are repeated iteratively until no further (or only marginal) increase in modularity is possible and the resulting partition corresponds to a local maximum of the modularity. An efficient implementation of Markov Stability optimisation across scales with the Louvain algorithm is provided by the \texttt{PyGenStability} python software~\cite{barahonaresearchPyGenStability2021}.

Optimisation of Markov Stability for different Markov scales $s$ then leads to a series of partitions $H(s)$. %
For small Markov scales, the Markov process can only explore  local neighbourhoods, which leads to a fine partition, whereas increasing the Markov scale widens the horizon of the Markov process so that larger areas of the graph are explored, which leads to coarser partitions~\cite{lambiotteRandomWalksMarkov2014}. Hence, the notion of a community as detected by Markov Stability analysis is strictly based on the spread of a diffusion on the network. In particular, Markov Stability analysis is able to identify non clique-like communities, which are likely present in heterogeneous mobility networks~\cite{schaubMarkovDynamicsZooming2012}.

\newpage 

\section{Pre-lockdown Facebook UK data: mobility graph, flow partitions, administrative regions and TTWAs}

\begin{figure}[H]
    \centering
    \includegraphics[width=.8\textwidth,trim={0 0 0 1cm},clip]{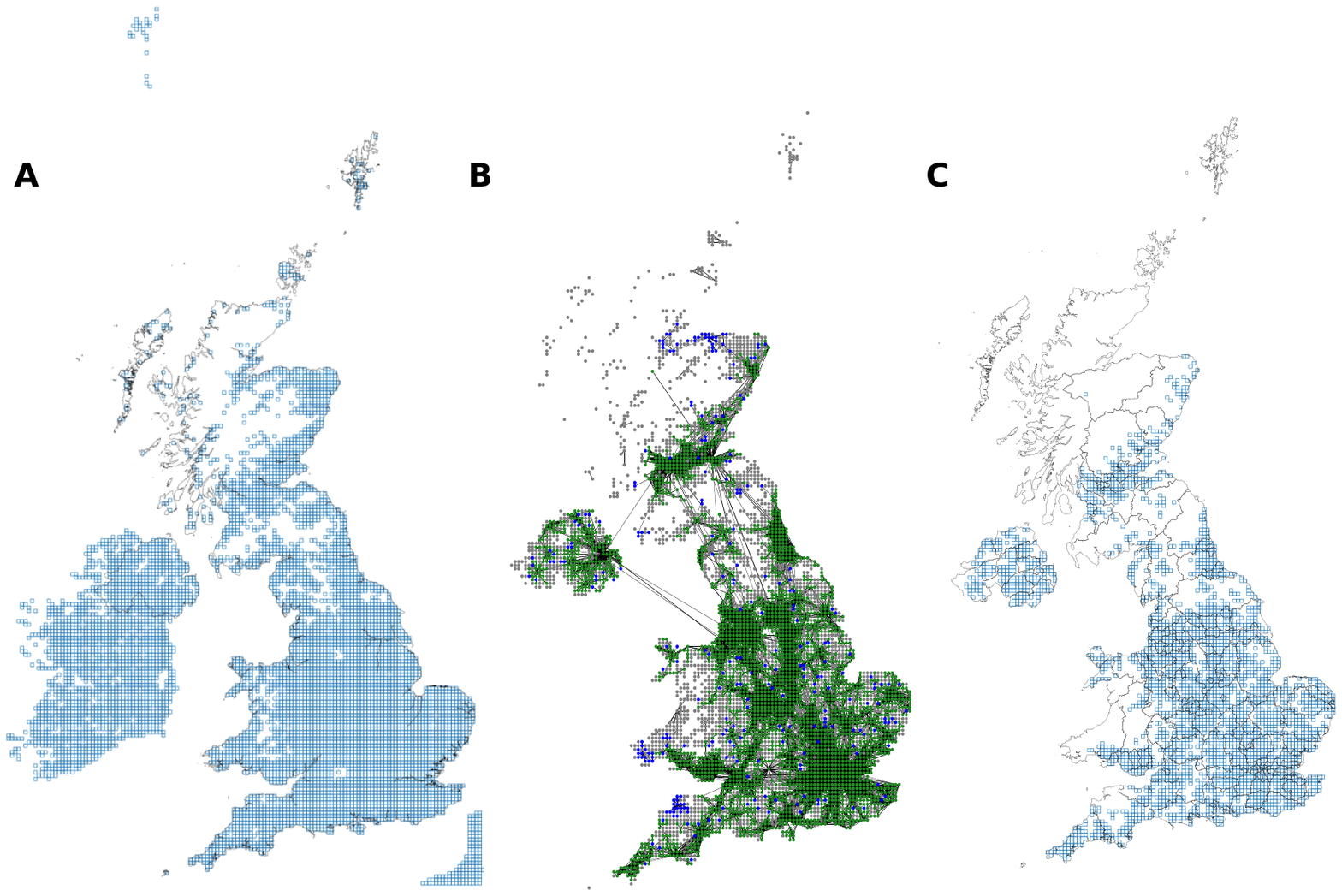}
   \caption[The Facebook Movement map pre-lockdown]{ \textbf{The Facebook Movement map pre-lockdown.} \textbf{A} The geographic 12-digit tiles for which movement data is available overlaid on the UK map. The tiles outside of the UK are not used. 
   \textbf{B} The directed mobility network, with nodes in the LSCC coloured in green; the remaining nodes in the LWCC in blue; and all other nodes in grey. Most of the grey nodes are unconnected to the rest of the network. \textbf{C} The geographic tiles in the LSCC are spread over most NUTS3 regions in the UK (borders shown by lines).}
   \label{S_fig:network_construction}
\end{figure}

\paragraph{Geographic tiles and mobility graph:}
As described in the main text, we used `Facebook Movement maps'\cite{maasFacebookDisasterMaps2019} to construct our baseline UK mobility graph. See Fig.~\ref{S_fig:network_construction} for some illustrative features of the data and graph.

\begin{figure}[H]
\centering
\includegraphics[width=.4\textwidth]{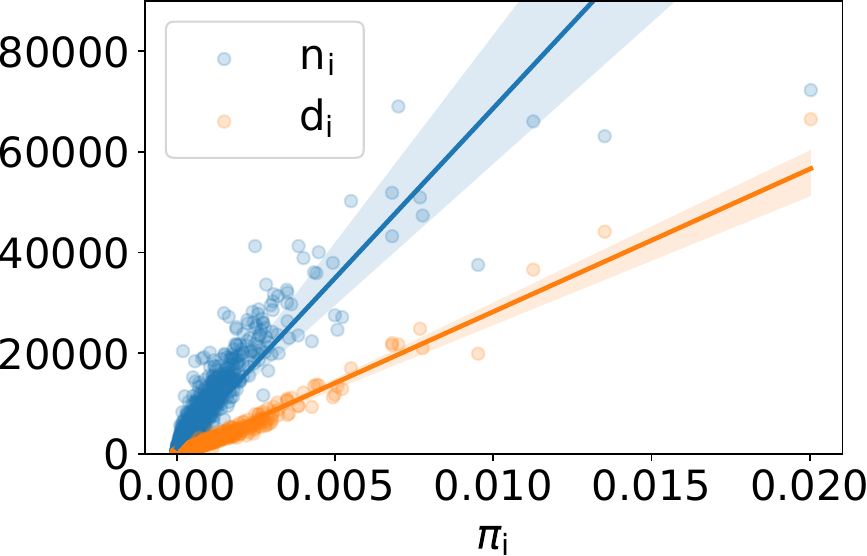}
\caption{\textbf{Correlation of stationary distribution $\boldsymbol{\pi}$ with out-degrees and intra-tile movement.} We observe that $\boldsymbol{\pi}$ correlates strongly with the out-degree $d_{\text{out}}$ ($R^2 = 0.97$) and slightly less so with the intra-tile movement $n_i$ within tile $i$ ($R^2 = 0.83$).}
\label{S_fig:Pi_correlations}
\end{figure}

\paragraph{Stationary distribution and other node centralities:}
Fig.~\ref{S_fig:Pi_correlations} analyses the correlation between the stationary distribution $\boldsymbol{\pi}$~\eqref{eq:stationary} and the out-degrees and intra-tile movement.

\begin{figure}[H]
\centering
\includegraphics[width=.5\textwidth]{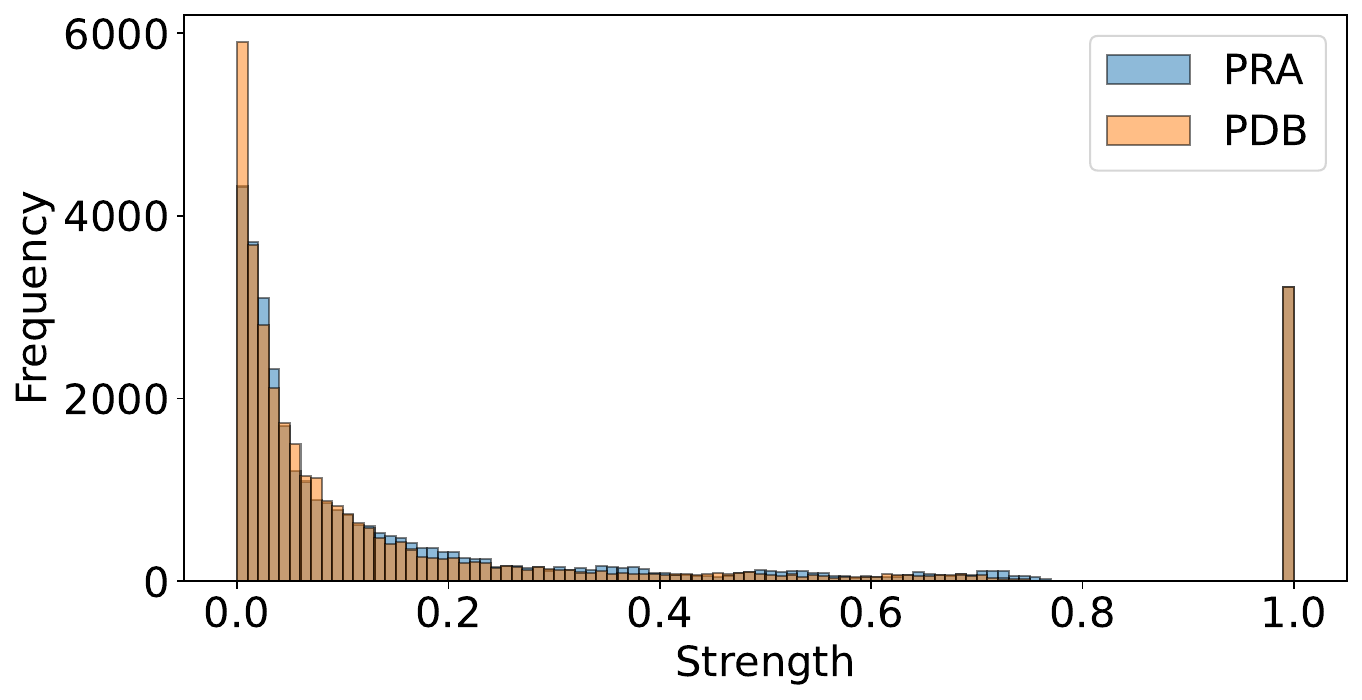}
\caption{\textbf{Asymmetry of baseline mobility graph as measured by the pairwise relative asymmetry and the pairwise detailed balance.}}
\label{S_fig:PRA_PDB_histogram}
\end{figure}

\paragraph{Detailed balance and asymmetry of graph diffusion at stationarity:} 
In analogy to the pairwise relative asymmetry (PRA) in the main text~\eqref{eq:PRA}, we define the \textit{pairwise detailed balance} (PDB) for each pair of tiles $ij$:
\begin{equation}\label{S_eq:PDB}
    0 \leq \text{PDB}_{ij} = \frac{|(\Pi M-\Pi M^T)_{ij}|}
    {(\Pi M+\Pi M^T)_{ij}} \leq 1.
\end{equation}
We find that the top 25\% of tile pairs have $\text{PDB} \geq 0.18$ (Fig.~\ref{S_fig:PRA_PDB_histogram}). Hence the diffusive process on the graph displays a reduced flow asymmetry at equilibrium, indicating the importance of (symmetric) commuting travel patterns.

\paragraph{MS partitions of the baseline mobility network:}
Table~\ref{S_tab:CommunityStats} provides additional statistics on the MS partitions.
\begin{table}[H]
\centering
\scalebox{1}{
\begin{tabular}[t]{|c| c| c c c|}
\hline
\multirow{2}{*}{Markov scale} & 
\multirowcell{2}{Number of \\ communities}
&  
\multicolumn{3}{c|}{Nodes per community} \\
& & Q1 & Median & Q3 \\
\hline
$s_1=-1.48$ & 201 & 	7 &	13 &	22 \\
$s_2=-0.38$ & 59 & 34 & 49 & 67 \\
$s_3=0.27$ & 32 & 67 & 87 & 116 \\
$s_4=0.71$ & 20 & 98 & 133 & 201 \\
$s_5=1.06$ & 16 &	107 & 168 &	274\\
$s_6=1.38$ & 13 &	126 & 263 & 313\\
$s_7=1.80$ & 9 & 246 & 274 & 376\\
$s_8=2.13$ & 7 & 260 & 319 & 560\\
$s_9=2.65$ & 6 & 253 & 365 & 835\\
\hline
\end{tabular}}\caption{\textbf{Statistics for MS partitions.} The number of communities and the three quartiles Q1, Q2 and Q3 for the number of nodes per community for the partitions at Markov scales $s_i$. %
}
\label{S_tab:CommunityStats}
\end{table}

\begin{figure}[htb!]
\centering
\includegraphics[width=.98\textwidth]{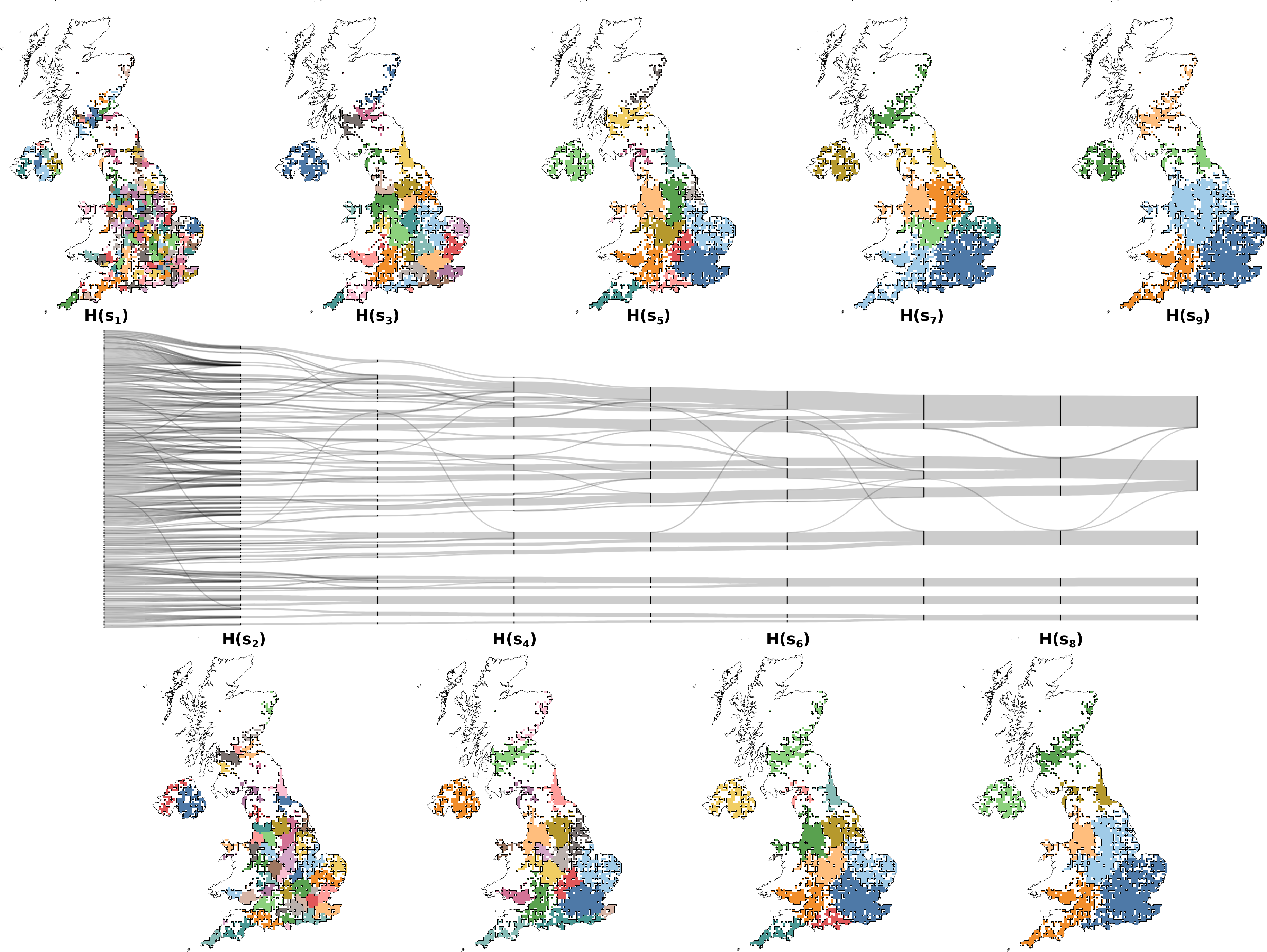}
\caption[Quasi-hierarchical structure of Markov Stability partitions]{\textbf{Quasi-hierarchical structure of Markov Stability partitions.} Small Markov scales correspond to fine partitions and large Markov scales to coarse partitions. One can observe a quasi-hierarchical structure in the Sankey diagram, which is not imposed by the algorithm but a feature of the data. 
Note that there is no consistency in the colours of communities in the maps across Markov scales.}
\label{S_fig:MS_Sankey}
\end{figure}

Fig.~\ref{S_fig:MS_Sankey} shows that the set of MS partitions $H(s_i)$ at Markov scales $s_i, i=1,\ldots, 9$ are quasi-hierarchical, a feature that is not imposed by the clustering algorithm but emerges as an intrinsic feature of the data.

\begin{figure}[H]
\centering
\includegraphics[width=.4\textwidth]{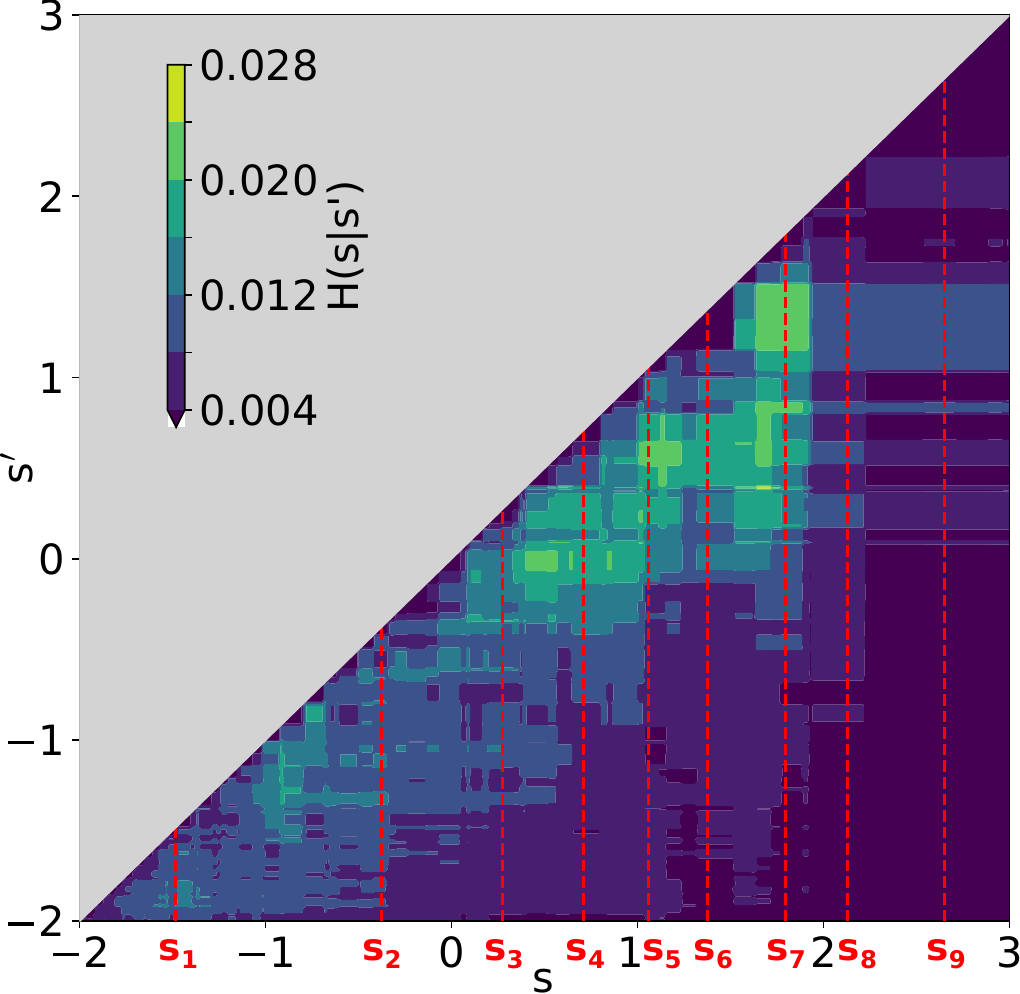}\caption{\textbf{Normalised Conditional Entropy of the sequence of MS partitions.} 
The robust MS scales are located in areas of low Normalised Conditional Entropy indicating a strong quasi-hierarchical structure. %
}
\label{S_fig:ConditionalEntropy}
\end{figure}

The intrinsic quasi-hierarchical nature of the set of partitions can be quantified using the Normalised Conditional Entropy~\cite{lambiotteLaplacianDynamicsMultiscale2009} $\hat{\mathcal{H}}(s|s^\prime)$ for two partitions $H(s)$ and $H(s')$ given by:
\begin{equation}\label{S_eq:ConditionalEntropy}
    0\le \hat{\mathcal{H}}(s|s^\prime) := \frac{\mathcal{H}(s|s^\prime)}{\log(N)}\le 1,
\end{equation}
where $\mathcal{H}(s|s^\prime)$ is the standard conditional entropy. This asymmetric quantity is a measure of hierarchy and vanishes if each community of the partition $H(s)$ is a union of communities in the partition $H(s')$. Fig.~\ref{S_fig:ConditionalEntropy} shows that the sequence of MS partitions displays a strong hierarchical structure with low values of $\hat{\mathcal{H}}(s|s^\prime)$.

\paragraph{NUTS administrative regions} Table~\ref{S_tab:NUTS_Stats} provides additional statistics for the NUTS regions. Following Brexit, International Territorial Level (ITL) geography hierarchy boundaries replace NUTS in the UK, but currently, ITL mirrors the NUTS European system. 
\begin{table}[H]
\centering
\scalebox{1}{
\begin{tabular}[t]{|c| c| c c c|}
\hline
\multirow{2}{*}{Level} & 
\multirowcell{2}{Number of \\ regions}
&  
\multicolumn{3}{c|}{Nodes per region} \\
& & Q1 & Median & Q3 \\
\hline
NUTS3 & 170 & 4 &	13 &	27\\
NUTS2 &	42 	& 40 &	72 &	102\\
NUTS1 &	12	&	230	& 254 &	324\\
\hline
\end{tabular}}\caption{\textbf{Statistics for NUTS regions (restricted to the geographic tiles in the baseline network).} 
Statistics for the three NUTS levels, including
the number of regions for each level and the three quartiles Q1, Q2 and Q3 for the number of nodes per region.  Note that NUTS regions where no movement was recorded are not present in the baseline data, hence the number of NUTS regions in the baseline network is smaller than the actual number of NUTS regions. %
}
\label{S_tab:NUTS_Stats}
\end{table}

\begin{figure}[H]
\centering
\includegraphics[width=0.8\textwidth]{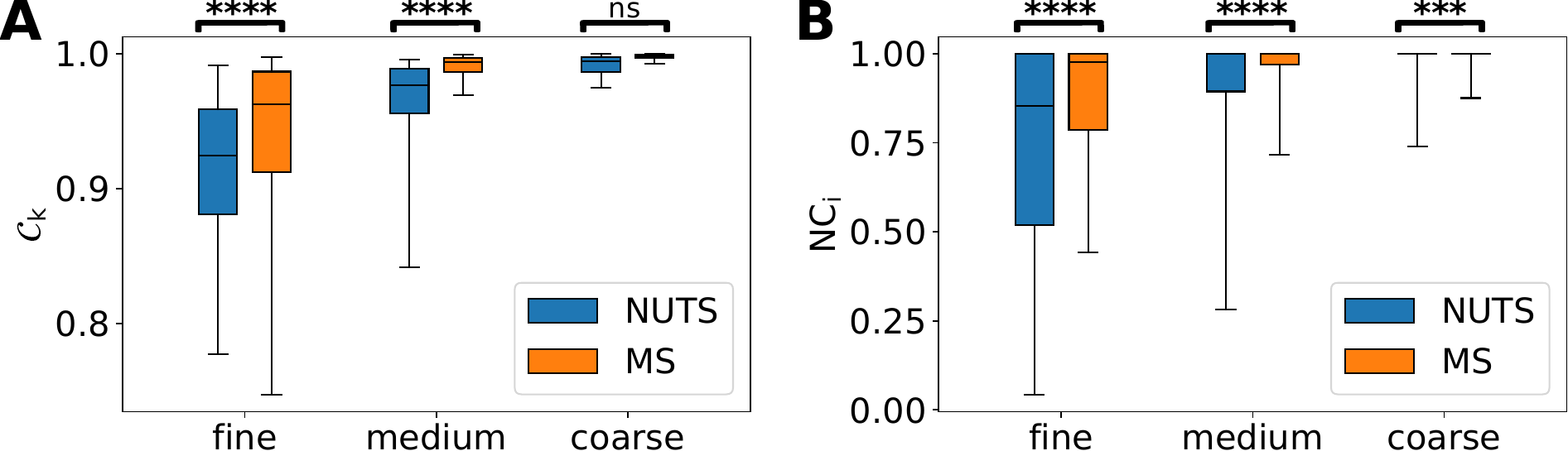}
\caption{\textbf{Comparison of Coverage and Nodal Containment between MS partitions and NUTS regions.} Fine (fine ($H(s_1)$ vs.\ NUTS3), medium ($H(s_2)$ vs.\ NUTS2) and coarse ($H(s_4)$ vs.\ NUTS1) levels of the MS partitions and NUTS regions are compared for \textbf{A} community Coverage~\eqref{eq:coverage_k}  and \textbf{B} Nodal Containment~\eqref{eq:NC_i}.  
Statistical significance is determined using the Mann-Whitney test (with **** indicating $p<0.0001$, and *** indicating $p<0.001$). MS partitions perform significantly better for each measure and have lower variability, indicating a better description of human mobility than the administrative NUTS regions.}
\label{S_fig:coverage_NC_significance}
\end{figure}

Fig.~\ref{S_fig:coverage_NC_significance} shows that the MS partitions better describe the patterns of human mobility both in terms of the community Coverage $\mathcal{C}_k$ and Nodal Containment $\text{NC}_i$.

\paragraph{Travel to Work Areas} As described in the main text, the TTWA division is a data-driven geography that divides the UK into 228 local labour markets. Our baseline network includes data for 197 of the 228 TTWAs in the UK and Table~\ref{S_tab:TTWA_Stats} provides additional statistics.
\begin{table}[H]
\centering
\scalebox{1}{
\begin{tabular}[t]{|c| c| c c c|}
\hline
\multirow{2}{*}{Level} & 
\multirowcell{2}{Number of \\ regions}
&  
\multicolumn{3}{c|}{Nodes per region} \\
& & Q1 & Median & Q3 \\
\hline
TTWA & 197 & 9 &	14 &	20\\
\hline
\end{tabular}}\caption{\textbf{Statistics for TTWAs (restricted to the geographic tiles in the baseline network).} 
Statistics for the TTWAs, including
the number of regions and the three quartiles Q1, Q2 and Q3 for the number of nodes per region.  Note that TTWAs, where no movement was recorded, are not present in the baseline data, hence the number of TTWAs in the baseline network is smaller than the actual number of TTWAs. %
}
\label{S_tab:TTWA_Stats}
\end{table}

\begin{figure}[H]
    \centering
    \includegraphics[width=0.7\textwidth]{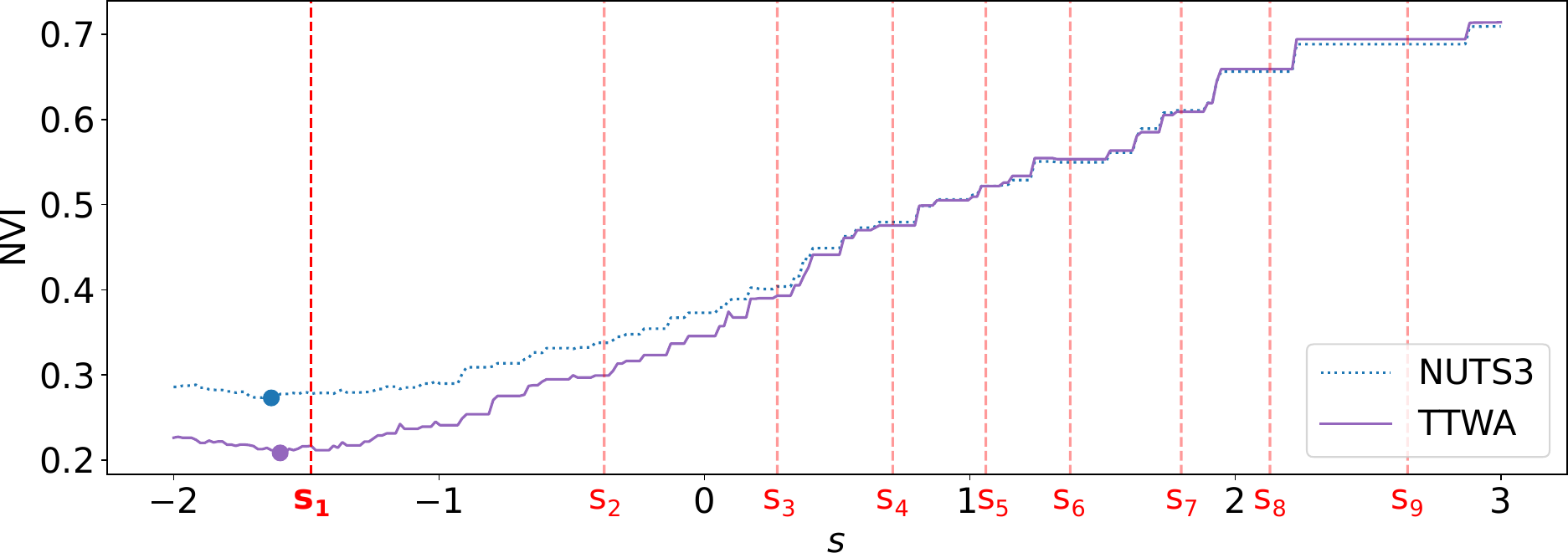}\caption{\textbf{\textit{A posteriori} comparison between Markov Stability (MS) partitions, the NUTS3 partition and the TTWA division.} The MS partitions across all scales are compared to the TTWA division and NUTS3. As indicated by the minima of the NVI, the TTWA division is closely similar to $H(s_1)$. For reference, we also include a comparison to the NUTS3 division and we observe that TTWA is more similar to the fine MS scale $H(s_1)$ as indicated by the lower NVI value.
    }
    \label{S_fig:comparison_NVI_TTWA}
\end{figure}

In Fig.~\ref{S_fig:comparison_NVI_TTWA} we use the Normalised Variation of Information (NVI)~\eqref{eq:NVI} to evaluate the similarity of the TTWA division to the MS partitions at all scales. The best match is close to the fine scale MS partition $H(s_1)$.

We also compare the TTWA division to the NUTS hierarchy of administrative regions, and find that the TTWAs are most similar to NUTS3 regions according to the lowest NVI value:
\begin{itemize}
    \item NVI(TTWA,NUTS1)= 0.55
    \item NVI(TTWA,NUTS2)= 0.39
    \item NVI(TTWA,NUTS3)= 0.28
\end{itemize}

\newpage
\section{Mobility response to lockdown restrictions}

\begin{figure}[H]
\centering
\includegraphics[width=0.5\textwidth]{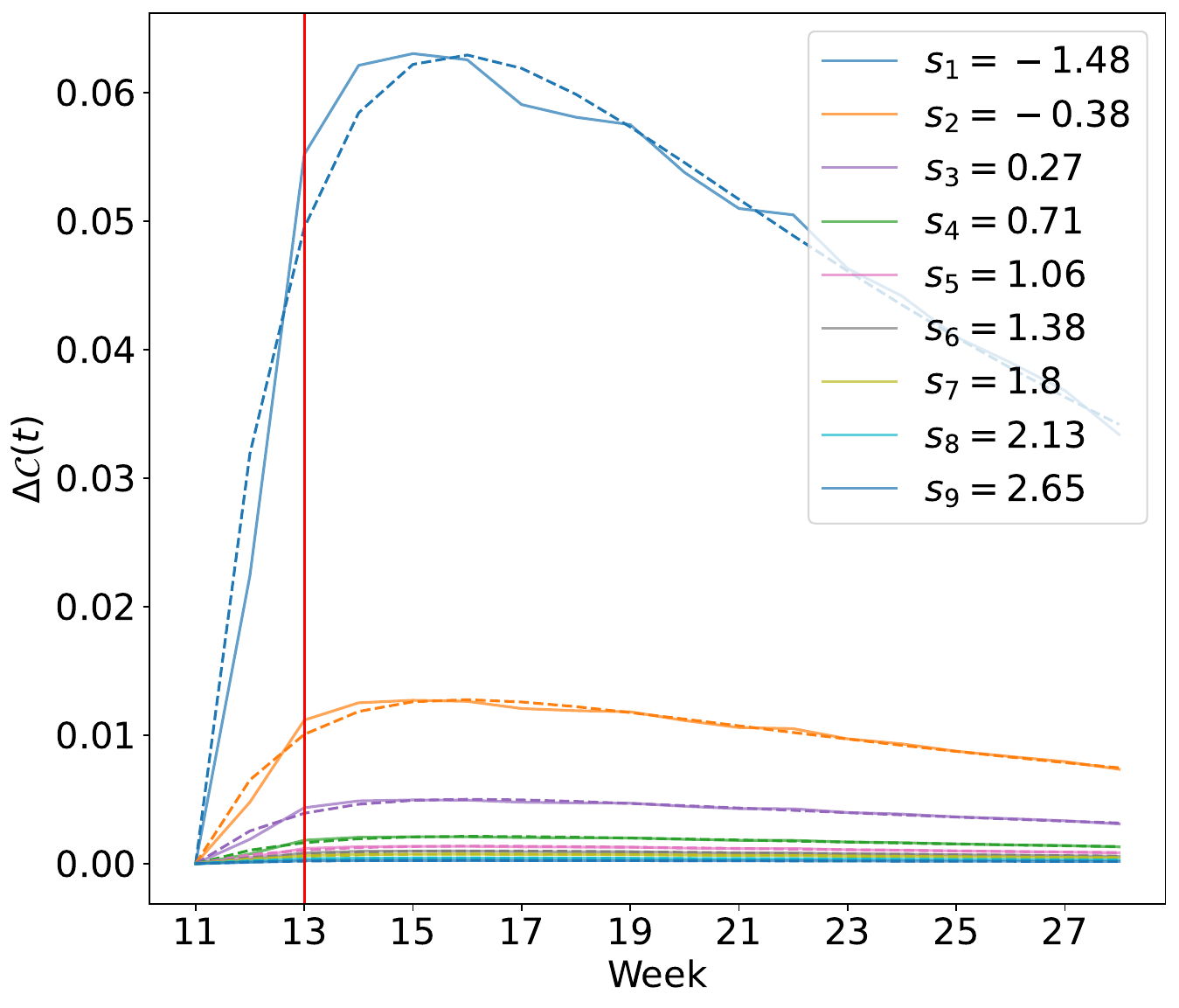}
\caption{\textbf{Temporal change of Coverage and Nodal Containment in response to lockdown restrictions.} The vertical red lines mark the official onset of the first lockdown in the UK on 24 March 2020.  %
\textbf{A} Relative change of Coverage $\mathcal{C}$~\eqref{eq:coverage_partition} of MS partitions at all scales.
The solid lines show the data and the dashed lines are the optimal fits of the activation response function~\eqref{eq:response_solution} to the weekly Coverage curves.
}
\label{S_fig:coverage_response_all_scales}
\end{figure}

Fig.~\ref{S_fig:coverage_response_all_scales} shows the relative change of the Coverage $\mathcal{C}$~\eqref{eq:coverage_partition} %
for the nine robust MS partitions, together with their fits to a linear response function with an exponentially decaying shock. 

Table~\ref{S_tab:Coverage_All_Response_Parameters} shows the parameters of the fits of $\Delta\mathcal{C}(t)$~\eqref{eq:response_solution} to the activation response function for all MS partitions.

\begin{table}[H]
\centering
\begin{tabular}[t]{l||c|c|c}
Markov scale &$\alpha$ (95\% CI) &$1/\beta$ (95\% CI) & $1/\lambda$ (95\% CI) \\
\hline
\hline
$s_1=-1.48$ & 0.042 (0.036--0.050) & 16.4 (12.5--21.5) & 2.0 (1.6--2.7) \\
$s_2=-0.38$ & 0.0086 (0.0074--0.0101) & 18.8 (14.6--24.3) & 1.92 (1.52--2.46)\\
$s_3=0.27$ & 0.0034 (0.0029--0.0040) & 21.8 (16.8--28.7) & 1.87 (1.49--2.38) \\
$s_4=0.71$ & 0.0014 (0.0012--0.0016) & 20.9 (15.7--28.0) & 1.98 (1.55--2.57) \\
$s_5=1.06$ & 0.00090 (0.00077--0.00106) & 21.4 (16.2--28.7) & 1.93 (1.52--2.49)\\
$s_6=1.38$ & 0.00063 (0.00053--0.00075) & 18.8 (13.8--25.5) & 2.07 (1.60--2.77)\\
$s_7=1.8$ & 0.00048 (0.00041--0.00057) & 20.9 (15.6--28.2) & 2.00 (1.57--2.61)\\
$s_8=2.13$ & 0.00030 (0.00026--0.00035) & 24.0 (18.7--31.6) & 1.80 (1.45--2.25)\\
$s_9=2.65$ & 0.00018 (0.00016--0.00021) & 26.5 (21.1--33.9) & 1.70 (1.41--2.06)\\
\hline
\end{tabular}
\caption{\textbf{Response parameters for Coverage.} Parameter estimates and 95\% Confidence Intervals for the shock amplitude $\alpha$, time scale of recovery $1/\beta$ and shock decay time $1/\lambda$ obtained from fitting the activation response function~\eqref{eq:response_solution} to $\Delta \mathcal{C}$.}  
\label{S_tab:Coverage_All_Response_Parameters}
\end{table}

\end{document}